\documentclass[useAMS,usenatbib]{mn2e}
\usepackage{graphicx}
\usepackage{fixltx2e}
\usepackage{subfigure}
\usepackage{times}
\usepackage{amsmath}
\usepackage{amsfonts}
\usepackage{url}
\usepackage{placeins}
\usepackage{natbib}
\usepackage{amssymb,amsmath}
\usepackage[normalem]{ulem}
\usepackage{soul}
\usepackage{lipsum}
\usepackage{longtable}
\usepackage{pdflscape}
\usepackage{amsmath}
\usepackage{multirow}
\usepackage{booktabs}

\usepackage{lipsum}
\begin{document}

\title[Small scale clustering]{Clustering on very small scales from a large sample of confirmed quasar pairs: Does quasar clustering track from Mpc to kpc scales?}
\author[Eftekharzadeh et al.]{S. Eftekharzadeh$^{1}$, A. D. Myers$^{1}$, J. F. Hennawi$^{2,3}$, S. G. Djorgovski$^{4}$, G. T. Richards$^{5}$,\newauthor A. A. Mahabal$^{6}$, M. J. Graham$^{4,7}$\\
\\
$^{1}$ Department of Physics and Astronomy, University of Wyoming, 1000 University Ave., Laramie, WY, 82071 \\
$^{2}$ Max-Planck-Institut fur Astronomie, Heidelberg, Germany \\
$^{3}$ Department of Physics, Broida Hall, University of California, Santa Barbara, CA 93106-9530\\
$^{4}$ Astronomy Department, California Institute of Technology, 1200 East California Boulevard, Pasadena, CA, 91125 \\
$^{5}$ School of Physics and Center for Relativistic Astrophysics, Georgia Institute of Technology, 837 State Street NW, Atlanta, GA
30332-043.\\
$^{6}$ Cahill Center for Astronomy and Astrophysics, California Institute of Technology, 1200 E California Blvd, MC 249-17, Pasadena CA, 91125 \\ 
$^{7}$ National Optical Astronomy Observatory, 950 N Cherry Avenue, Tucson, AZ 
85719}

\maketitle

\begin{abstract}

We present the most precise estimate to date of the clustering of quasars on very 
small scales, based on a sample of 47 binary quasars with magnitudes of $g<20.85$ and proper transverse separations of $\sim 25\,h^{-1}$\,kpc.
Our sample of binary quasars, which is about 6 times larger than any previous spectroscopically confirmed
sample on these scales, is targeted using a Kernel Density Estimation technique (KDE) 
applied to Sloan Digital Sky Survey (SDSS) imaging over most of the SDSS area. 
Our sample is ``complete" in that all of the KDE target pairs with $17.0 \lesssim R \lesssim 36.2\,h^{-1}$\,kpc in
our area of interest have been spectroscopically confirmed from a combination of 
previous surveys and our own long-slit observational campaign. We catalogue 230 candidate quasar pairs with angular separations of $<8\arcsec$, 
from which our binary quasars were identified. We determine the 
projected correlation function of quasars ($\bar W_{\rm p}$) in four bins of proper 
transverse scale over the range $17.0 \lesssim R \lesssim 36.2\,h^{-1}$\,kpc. 
The implied small-scale quasar clustering amplitude from the projected correlation function, integrated across our 
entire redshift range, is $A=24.1\pm3.6$ at $\sim 26.6 ~h^{-1}$\,kpc.
Our sample is the first spectroscopically confirmed sample 
of quasar pairs that is sufficiently large to study how quasar clustering evolves with redshift at $\sim 25 ~h^{-1}$ kpc.
We find that empirical descriptions of how quasar clustering
evolves with redshift at $\sim 25 ~h^{-1}$ Mpc also adequately describe the evolution of quasar
clustering at $\sim 25 ~h^{-1}$ kpc.
\end{abstract}



\section{Introduction}\label{intro}

Quasars, like galaxies, are biased tracers of the underlying dark matter distribution
\citep[e.g.,][]{col89,bw02}. Many models invoke galaxy mergers as the mechanism
for triggering quasar activity, although the necessity of such a mechanism is still 
debated \citep[e.g.,][]{coi07,pad09,gr11}. Certainly, though, 
structure formation models can reproduce quasar demographics
under the assumption that quasar 
activity is triggered by mergers \citep[e.g.,][]{wy05}. 
The peaks of the density field in which quasars reside might have 
been particularly strongly clustered, given that mergers are more frequent in denser 
environments \citep[e.g.,][]{kai84,lc93,dm05,hop08} and that density signals from mergers can persist
on timescales similar to the lifetime of quasars \citep{Wet09}.
Quasar clustering measurements therefore offer a tool by which to understand the physical 
processes that trigger quasar activity. 
 The ongoing attempts to conduct such investigations
 become more challenging at higher luminosities \citep[e.g.,][]{el11,el13,ji16}.   

Surveys such as the Sloan Digital Sky Survey (SDSS), 
increased the sample size and number density of quasars in a large volume of 
space, substantially improving measurements of quasar clustering on large or ``two-halo" scales 
\citep[e.g.,][]{po04,cr05,my06,po06,my07,she07,she09a,ro09,wh12,ef15}. 
Measuring quasar clustering on small scales, however, is 
more challenging for several reasons. First, quasars with small angular 
separations ($< 60\arcsec$) are simply rare. Second, surveys that use fiber-fed multi-object spectrographs, such as the SDSS, 
prevent fibers from colliding by never placing two fibers closer than 
about 60\arcsec\,on a single plate \citep[]{bl03,Daw13}. Third, finding rare quasar pairs without exploiting large
surveys typically requires many individual long-slit observations of pairs of candidates, which is time-consuming.

The first small-scale quasar pairs were often discovered by chance in dedicated fields, or 
during long-slit surveys for gravitationally lensed quasars\footnote{SDSS J1637+2636AB was the first binary QSO discovered, but originally misinterpreted as a lens \citep{dj84}. Due to this misinterpretation, the quasar pair PKS 1145-071 was initially known as the first binary quasar \citep{dj87}.} \citep[e.g.,][]{sw78,we82,cr88,he89,me89,sch94,ha96,fan99,koch99,mort99,sch00,gr02,mi04,pin06,mg16}. 
 Although the search for high redshift quasar pairs dates back to individual discoveries of quasars at $z\sim4$  \citep[e.g., ][]{cr88,mc88,me90,dj91,sch94,he98,zh01},
with the development of photometric selection algorithms to build homogeneous 
sets of quasar candidates from large imaging surveys \citep[e.g.,][]{ric04,ric09}, it became possible to conduct more
homogeneous searches by prioritizing highly-probable close quasar pairs and following them up with long-slit spectroscopic 
surveys \citep[e.g.,][]{he06a,my07b,my08}. These surveys focused on quasar pairs separated
by less than 2000\,${\rm km}\,{{\rm s}^{-1}}$ in redshift-space in order to measure small-scale clustering, denoting such pairs
``binary quasars,'' a term that has appeared for decades in the literature \citep[e.g.][]{mu98}. \citet{he06a}
elucidate the specific use of a velocity range of $|\Delta v|< 2000~{\rm km}\,{{\rm s}^{-1}}$ as being wide enough to cover the most prominent sources of redshift uncertainty for quasars.
In particular, peculiar velocities of up to $500~{\rm km}\,{{\rm s}^{-1}}$ in dense environments and blueshifted broad lines of 
up to $1500~{\rm km}\,{{\rm s}^{-1}}$ \citep{rich02a,he06a}.

In tandem with similar homogeneous searches for gravitational lenses
\citep{og06,og08,in08,in10,og12,in12} work on binary quasars has driven measurements of quasar clustering on
very small scales down below even a few hundred kiloparsecs ($\sim10\arcsec$ or lower). For example,
\citet{ko12} took advantage of the 
Sloan Digital Sky Survey Quasar Lens Search \citep{in12}, 
to measure the quasar correlation function down to $\sim 10 \,h^{-1}$\,kpc.

In this paper, we continue in the vein of \citet{he06a}, \cite{my07b} and \citet{my08}. 
We identify high-probability candidate close quasar pairs from  
a homogeneous catalogue of candidates and follow them up with confirming spectroscopy. 
Our target sample is drawn from quasar candidates
selected using Kernel Density Estimation (KDE) by \citet{ric09}. This ``KDE'' sample is
not only large, it is pure\footnote{$>90$\% of KDE candidates at $0.4\,\lesssim z \lesssim 2.3$ are expected to be quasars}, so presents an efficient parent sample to mine for binary quasars.
The sample of 47 confirmed binary quasars that we will discuss in this paper is complete for angular separations of $2.9\arcsec <\theta< 6.3\arcsec$ and redshifts of $0.43<z<2.26$.
Note that we will use the term ``complete'' here, to refer to 100\% confirmation of whether or not all of our candidate target pairs are a binary quasar. We do not
mean complete in the sense of also capturing quasars that are {\em not in} the KDE catalogue.
Our sample improves on previous work in being over five times larger than previous samples of binaries 
on the range of scales that we cover ($\sim 20$ -- $40\,h^{-1}$\,kpc).
The $>2\times$ more precise correlation function that we calculate over proper separations at $<50 \,h^{-1}$\,kpc is therefore
the tightest constraint on quasar clustering to date on scales of a few tens of kiloparsecs.

This paper is structured as follows: The data are introduced in $\S $\ref{dat}, 
and our methodology for measuring and modeling clustering is 
discussed in $\S $\ref{cls}. $\S $\ref{dis} is dedicated 
to the interpretation of our clustering results, before we summarize our work in $\S$\ref{sum}.
We adopt a $\Lambda$CDM cosmological model 
with $\Omega_{m}=0.307$, $\Omega_{\Lambda}=0.693$, $h=0.677$ consistent with 
\citet{planck15}. All distances quoted throughout the paper are in proper coordinates unless mentioned otherwise. We convert measurements from the literature to 
proper coordinates prior to comparing such measurements to our results. We use ``cMpc'' and ``ckpc'' to denote comoving distance units when we compare our measurements in proper coordinates to 
correlation lengths in comoving coordinates that have been derived from Mpc-scale clustering measurements.
In our chosen cosmology, an angular separation of $1\arcsec$ at $z=1.5$ corresponds to a proper separation of $ 5.9 ~h^{-1}$\,kpc.

\section{Data}
\label{dat}
\subsection{Identification of new quasar pairs}
Our starting sample consists of 1{,}172{,}157 high-probability candidate quasars 
identified by \citet{ric09} using Kernel Density Estimation \citep[henceforth KDE; see also][]{ric04}. 
\citet{ric09} applied the KDE technique to a test sample consisting of all point sources in SDSS Data Release 6 \citep[DR6; ][]{admc08} imaging down to a
limiting magnitude of $i=21.3$. These test data were 
labeled as a ``star'' or a ``quasar'' using a non-parametric Bayesian classifier, based on their position in $ugriz$ colour space. The PSF-magnitudes of the sources were extinction-corrected based on the \citet{schl98} dust maps.
The density of the ``quasar'' and 
``star'' colour space was established by applying the
KDE technique to ``training'' samples of stars and quasars. The ``stars'' training set resembled a randomly drawn subset of the test set. The ``quasars'' training set consisted of
spectroscopically confirmed SDSS quasars \citep{sch07} largely limited to $i<19.1$ at $z<3$ and $i< 20.2$ at $z>3$.
At higher redshift, the quasar training sample was supplemented by quasars from the AAOmega-UKIDSS-SDSS (AUS) QSO survey 
(Croom et al., in prep) and from \citet{fan06}.
Given that the position of the quasar locus
in colour space relative to that of the stellar locus changes significantly with 
redshift, \citet{ric09} conducted a redshift-and-colour-based sub-classification 
in four narrower ranges of; low redshift ($z \le 2.2$); intermediate redshift ($2.2 \le z \le 3.5$); high redshift ($z \ge 3.5$); and also UV-excess (UVX), based on $u-g$ colour. 
High-probability quasars classified in these ranges
are denoted {\tt lowzts==1}, {\tt midzts==1}, {\tt hizts==1} or {\tt uvxts==1}, respectively \citep[see Table 2 of][]{ric09}.

From the initial KDE sample of 1{,}172{,}157 candidate quasars, we sub-selected candidates that are brighter than 20.85 in (Galactic-extinction-corrected) $g$-band and
are categorized as {\tt hizts==1} OR {\tt uvxts==1}. We further restricted this subsample to the $70^\circ < {\rm RA} < 300^{\circ}$ region of the DR6 imaging footprint, resulting in
a total of 369{,}559 quasar candidates.
We will hereafter refer to these 369{,}559 candidates as our ``parent sample''. We cross-matched the candidates by
angular separation and identified 230 candidate pairs with  separations of $2.8\arcsec < \theta < 8\arcsec$. 
Here, 
the upper limit is chosen to correspond to a few hundred kpc for likely quasar redshifts. The lower limit is chosen to match roughly twice the seeing of the SDSS imaging data \citep[e.g.\ see Figure\ 4 of][]{DR1} in order
to protect against sources that are merged in SDSS imaging. 

To determine which of our candidate pairs had already been identified as quasars, 
we used a radius of 1\arcsec\ to cross-match our parent sample with previously known, spectroscopically confirmed, visually inspected, quasars. 
These ``known'' quasars were drawn from programs
conducted to identify quasar pairs \citep{he06b,he06a,my08,hen10,hp13,pr13b,pr14} and gravitational lenses \citep{og08,og12,in08,in12} as well as SDSS 
Data Release 7 \citep[DR7;][]{sch10} and Data Release 12 \citep[DR12;][henceforth DR12Q]{par16}. In particular, DR12Q
includes objects from an SDSS ancillary program designed specifically to target some of our candidate quasars\footnote{\url{http://www.sdss.org/dr12/algorithms/ancillary/boss/smallscaleqso/}}.
We then identified candidate quasar pairs that did not have {\em both} members of the pair previously spectroscopically confirmed, and 
reserved such ($2.8\arcsec < \theta < 8\arcsec$) pairs for further spectroscopic confirmation. 

\begin{figure*}
\includegraphics[angle=0,scale=0.4]{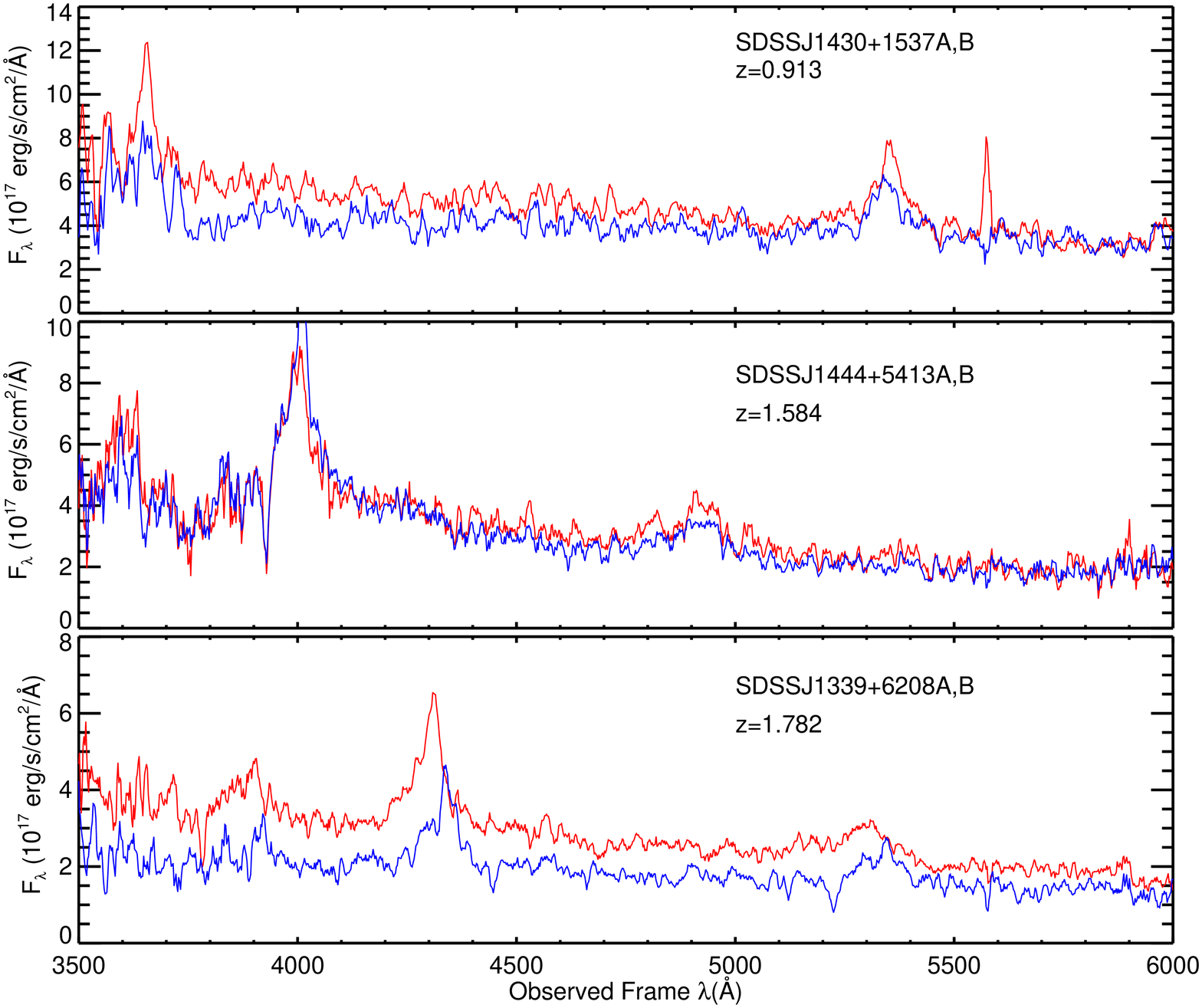}
\caption{Three example spectra of confirmed binary quasars. 
The spectra are smoothed by 5 pixels to aid visualization. }
\label{fig1:rl}
\end{figure*}

\begin{figure*}
\includegraphics[angle=0,scale=0.4]{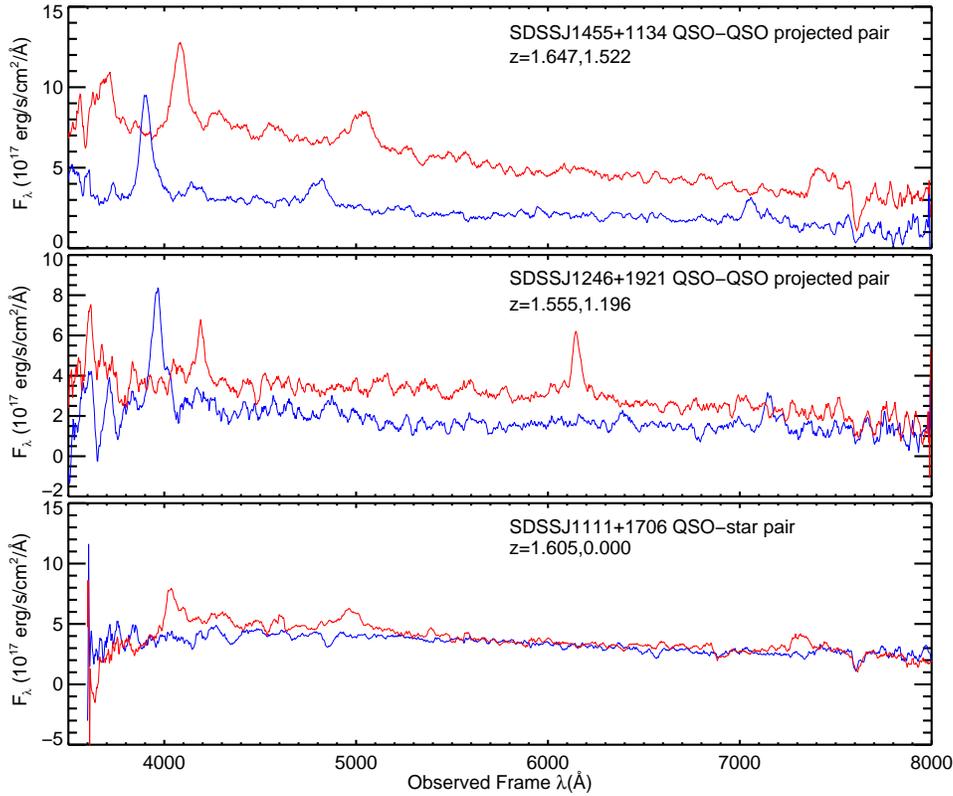}
\caption{Three example spectra of quasar pairs that are not binaries. Such pairs 
can be two quasars that are aligned along the line of sight but have different redshifts,
star-quasar pairs, or star-star pairs.
}
\label{fig2:pj}
\end{figure*}

Long-slit spectroscopy of these selected candidate quasar pairs 
was conducted on a range of facilities outlined in Table \ref{ins}, with the slit oriented to observe both quasars simultaneously.
The spectra were reduced and calibrated using the XIDL package\footnote{\url{http://www.ucolick.org/~xavier/IDL/index.html}}. Figures \ref{fig1:rl} and \ref{fig2:pj} show three examples of reduced spectra of our quasar pairs. 
Table \ref{full} contains the full list of 230 candidate quasar pairs drawn from our parent sample, together with available spectroscopic confirmations and redshifts from our own and previous campaigns.

\begin{figure}
\begin{center}

\centering
\includegraphics[angle=0,scale=0.5]{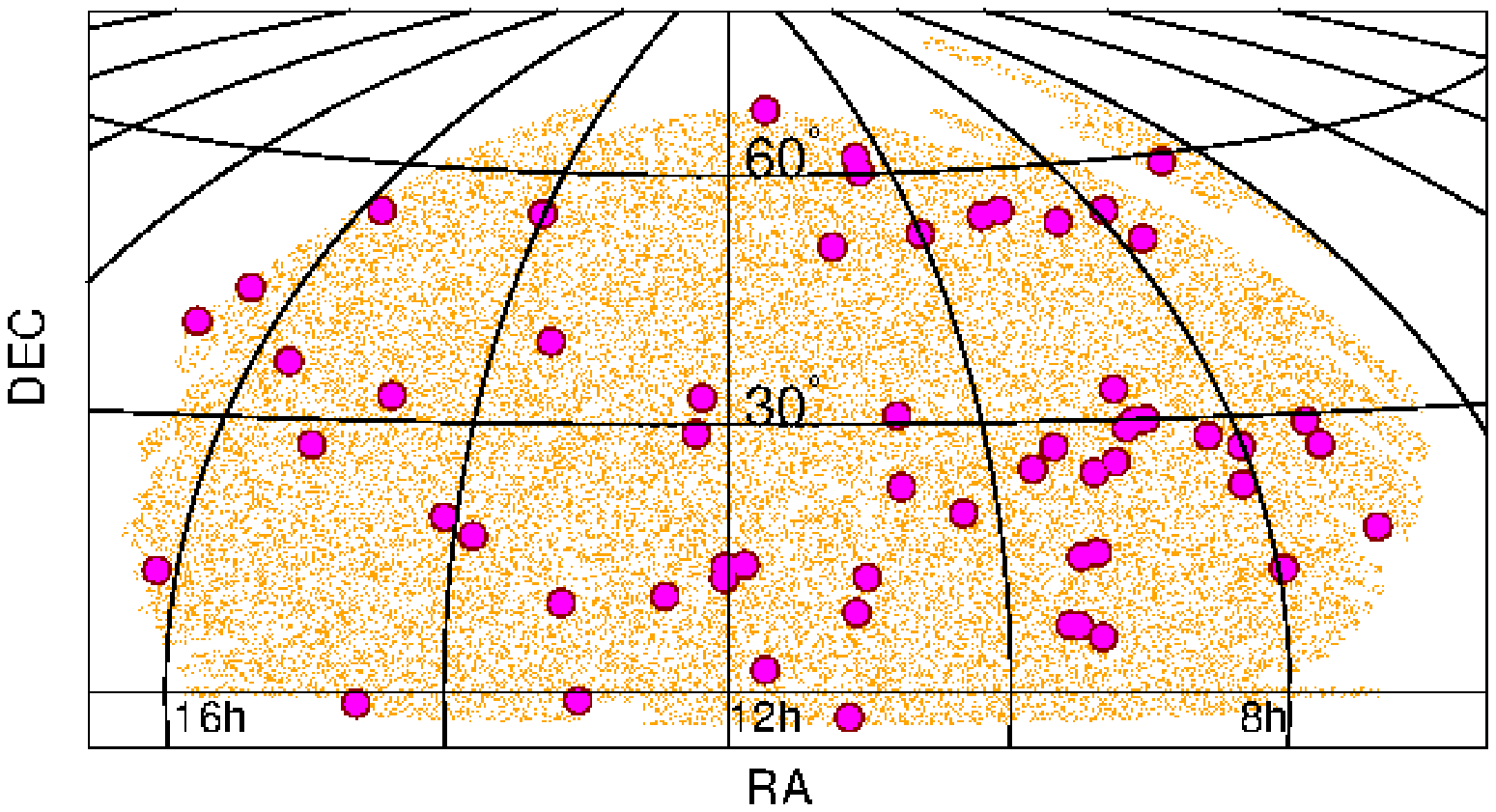}
\caption{Coordinates of the 369{,}559 quasar candidates in our parent sample (orange) in Aitoff projection.
The filled circles depict the 58 spectroscopically confirmed binaries in our sample of relevant pairs. The data have been cut
to the NGC imaging footprint of SDSS DR6, our main area of focus.
}
\centering
\label{fig3:aitoff}
\end{center}
\end{figure}
\subsection{A KDE-complete sample of binary quasars}
\label{sec:KDEcomp}

Our goal is to characterize quasar clustering on very small scales using a statistically uniform sample 
of quasars that are proximate to each other (so-called ``binary'' quasars).
Following \cite{he06a}.
We designate pairs of quasars that do not meet this criterion to be ``projected pairs.'' Over the course of our campaign to date, we have obtained definitive classifications for a close to complete sample of KDE-selected
candidates on angular scales
of $2.9\arcsec\,\le\,\theta\,\le\,7.7\arcsec$, which correspond to proper scales\footnote{We use the angular separation and confirmed redshift of the {\em brighter} member in each of our relevant pairs to calculate the proper transverse separation between members of the pair.}  of roughly $15\,\le\,R\,\le\,40\,h^{-1}$\,kpc over the main redshift range of our sample. 
The resulting sample consists of 169 candidate quasar pairs, which we will refer to as our sample of  ``relevant pairs''.
Note that good spectroscopy of {\em both} candidates is not required to ``definitively classify'' a pair as {\em not} a binary quasar. For instance, if one of a pair of 
objects is categorically identified as a star or a galaxy, then that pair is a non-binary, and we classify it as a ``projected pair.'' Further, if a known 
quasar at a redshift of $z$ has a companion for which we have a spectrum that is of sufficiently high quality that 
we should certainly have identified broad emission lines corresponding to $z$, then we also classify that pair as a projected pair. 
Note that we do not consider confirmed quasar pairs to be ``binary'' even if their velocity separation is only slightly larger than 2000\,km\,s$^{-1}$. In addition, we
removed one pair\footnote{SDSS\,J1336+2737 with a separation of 5.41\arcsec} from our ``relevant pair'' sample that consisted of two high signal-to-noise but featureless (``continuum'') sources. Even if this pair is a binary
quasar, we would have no way to assign it a redshift. 

Table~\ref{tab:169pairs} records the nature of our total of 169 relevant pairs, including their ultimate classification as a binary quasar, a projected pair, a pair for which there is insufficient
information to characterize it, or a gravitational lens\footnote{We designate binary quasars as gravitational lenses if they 
are convincingly argued to be lenses in the literature. }. The distribution on the sky of the binary quasars in our sample of relevant pairs is shown in Figure \ref{fig3:aitoff},
against a background of all of the KDE-selected candidates in our parent sample. Our follow-up spectroscopy of candidate pairs, provided 126 {\em new} sets of observations of 
candidate quasar pairs that have separations of less than $7.7 \arcsec$.
Of these 126 newly characterized pairs, we confirmed 53 to be binary quasars.  \citet{ric09} used clustering analyses to estimate that the KDE selection algorithm 
is 92.7\% efficient for sources with {\tt hizts==1} OR {\tt uvxts==1}. If we designate as ``stars'' those objects in our sample that do not have a sufficiently good spectrum to
classify the object\footnote{a reasonable assumption, given that quasars are much easier to classify at low-signal-to-noise as compared to stars.}, then we find that
out of the 338 sets of candidate quasars in our sample of 169 relevant pairs, we confirm 309 to be quasars. 
This is in excellent agreement with an efficiency of $\sim$92.5\% for the KDE catalogue.

Typically, clustering studies construct a random catalogue, or otherwise analytically correct for sources of incompleteness that arise when targeting quasars \citep[e.g.\ Eqn.\ 17 of][]{he06a}.
To circumvent incompleteness corrections when conducting clustering analyses, we instead construct a sample of pairs that we have categorically identified as either a binary quasar  or not. 
We will henceforth refer to this subset of pairs as our ``KDE-complete'' sample. Binary quasars in the KDE-complete redshift and proper scale ranges can be used for clustering analyses without correcting for
incompleteness in our spectroscopic campaign (because, by definition, this range is 100\% complete to possible binary quasars in our parent sample).
The outer limits of our KDE-complete sample in redshift and proper transverse scale are defined by the ranges at which there exist quasar pairs that we cannot categorically classify as a binary or not.
Typically, this is because the spectroscopic information for either one or both members of the pair does not exist. Note that there are cases where spectroscopic confirmation of only one member of a pair
is sufficient to include that pair in the KDE-complete sample. Most obviously, as also noted above, pairs that include one non-quasar have sufficient information to be included in the KDE-complete sample. In addition, though,
pairs that include one confirmed quasar with a (spectroscopic) redshift that would categorically place it outside of the proper-scale range of interest can also be use to define the KDE-complete ranges, regardless of whether
such a quasar's companion has itself been spectroscopically confirmed.

\begin{figure*}
\includegraphics[angle=0,scale=0.4]{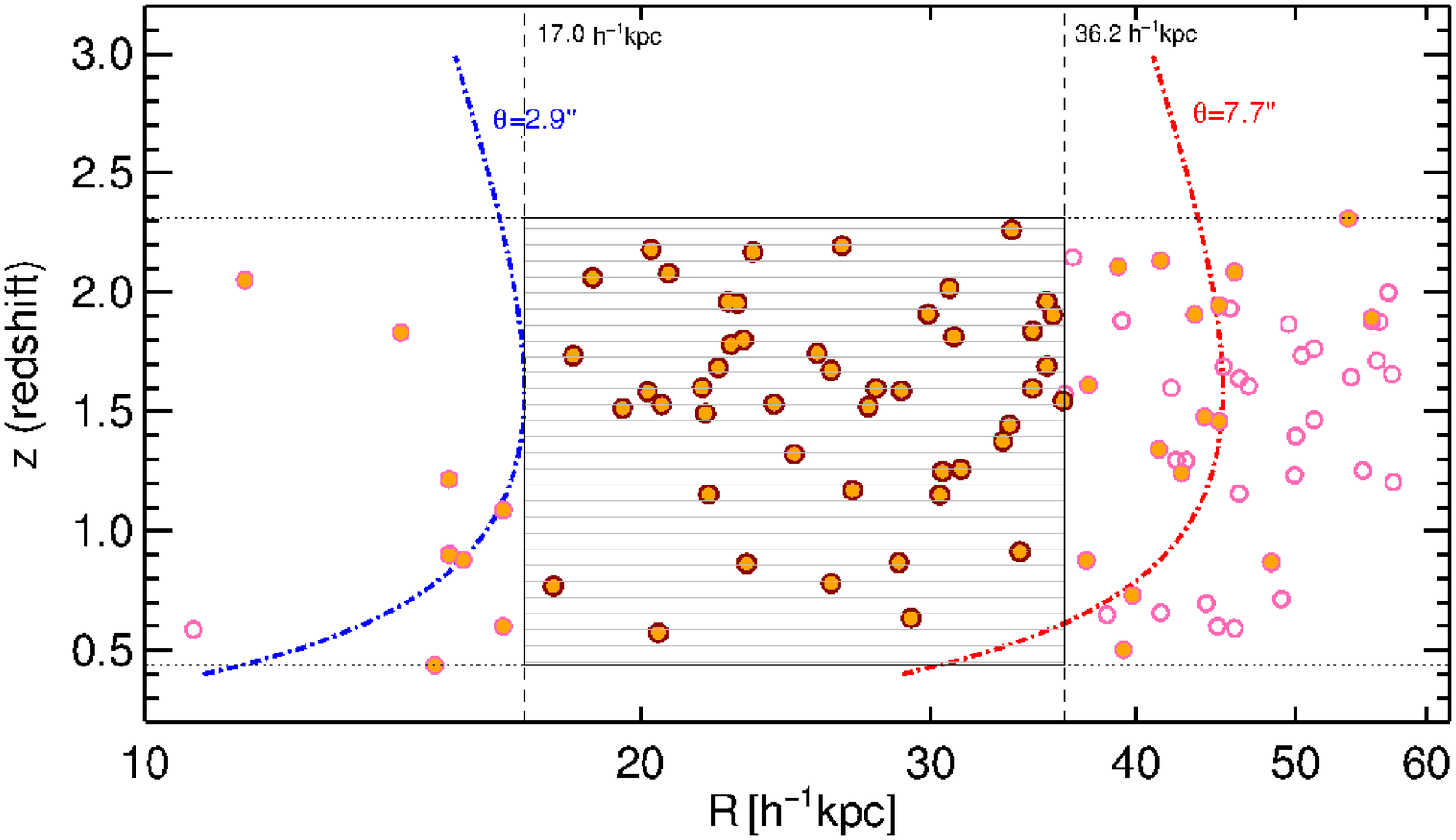}
\caption{Redshift and proper transverse separation range probed by the binary 
quasars in our ``KDE-complete'' sample. Filled circles represent 
spectroscopically confirmed binary quasars. Open circles represent binary quasars
for which spectroscopic 
information exists for only one of the members of the pair. The 
dotted lines depict transverse separations corresponding to 
$2.9\arcsec\,\le \theta \le 7.7\arcsec$, the angular range of our
169 relevant candidate quasar pairs. The extent of our ``KDE-complete'' 
sample of 47 binaries is depicted by a 
grey box that is
limited by either the angular extent of our sample of relevant pairs, 
or by quasar pairs that we cannot currently confirm or reject as a binary based on spectroscopy. 
}
\label{fig4:Rz}
\end{figure*}

Figure \ref{fig4:Rz} shows the redshift and proper transverse separation ranges for binary quasars in our KDE-complete sample.
The dotted lines show the transverse separations corresponding to the $2.9\arcsec \le \theta \le 7.7\arcsec$ angular range of our
169 relevant pairs.
The extent of the KDE-complete sample of binaries both in redshift and in transverse separation is depicted by a 
grey box. This box is limited by either the angular extent of our sample of relevant pairs, 
or by quasar pairs that we cannot currently confirm or reject as a binary based on the spectroscopic information to hand
(depicted by open circles in Figure \ref{fig4:Rz}). 
The ranges of redshift and proper separation that define the limits of the KDE-complete sample (i.e.\ the edges of the grey box in Figure \ref{fig4:Rz}) are $0.44 
\le z \le 2.31$ and $17.0\le R \le 36.2 \,h^{-1}$\, kpc. Table \ref{tab47} lists the sample
of 47 binary quasars that define our KDE-complete sample. Figure~\ref{fig4:Rz} also illustrates that we only consider a small fraction of space with $\theta < 3\arcsec$, in
keeping with the arguments in \citet{Pin03} and \citet{he06a} that sources with $\theta < 3\arcsec$ can appear blended in SDSS imaging.

\begin{table*}

\begin{tabular}{ccccclc}
\hline
\hline
                                                                                                                                      
Telescope &     Instrument     &       Spectrograph     &  Spectral     &   FWHM     & Dates & Reference\\ 
 & & Type & Coverage (\AA) & & & \\ \hline

Mayall 4-m  &   Ritchey-Chr\'{e}tien Spectrograph (RC)  &   Single  &  3600 -- 9200       &  325&       9-11 Feb., 7-10 Jun. 2008 & (1) \\
Palomar 200 inch   &   Double Spectrograph (DBSP)    &   Double     &   3100 -- 9300    &      900/550         &  28-29 Feb., 2-5 Apr. \& 4-6 May 2008  & (2) \\
Palomar 200 inch   &   Double Spectrograph      &   Double     &  3100 -- 9300   &        900/550       & 24 Feb., 30 Mar., 27 Apr.  \& 17 Jun. 2009  &  (2) \\
Palomar 200 inch   &   Double Spectrograph    &   Double     &   3100 -- 9300    &     900/550          &  7-10 Nov. 2010  & (2) \\
Palomar 200 inch   &   Double Spectrograph         &  Double  & 3100 -- 9300     &   900/550    & 2-3 Mar. 2011 & (2) \\

\hline
\end{tabular}

\caption{Summary of the follow-up spectroscopic campaign for a complete 
subsample of the KDE-selected quasar candidates. The {\em Reference} column 
refers to (1) \citet{shei02}; (2) \citet{og83}. }\label{ins}
\end{table*}

\begin{table}
\begin{tabular}{lc}
\hline
\hline
Category  & \# of pairs \\

\hline
Confirmed binaries & 58\\
Confirmed lenses &  5\\
Confirmed quasar pairs (non-binaries) & 77 \\
Pairs with at least one confirmed non-quasar member& 8\\
Pairs with at least one unknown member & 21\\

\hline
\end{tabular}
\caption{Classification of all 169 ``relevant pairs'' in our sample (with 
$g<20.85$ and $2.9\arcsec < \Delta \theta < 7.7\arcsec$). ``Confirmed binaries''  meet the classification of a binary quasar
for the purposes of this paper ($|\Delta v_{||}| < 2000 \,\rm km\,\rm s^{-1}$ and not
otherwise identified as a gravitational lens); ``Confirmed quasar pairs'' denotes
pairs for which we have spectroscopic information for both members of the candidate pair and
that have $|\Delta v_{||}| \geq 2000 \,\rm km\,\rm s^{-1}$.}\label{tab:169pairs}
\end{table}

\section{Methodology}

\subsection{Estimating the small-scale clustering of quasars}
\label{cls}

We measure the correlation function in proper coordinates, projected across a redshift window
of $< 2000\,\rm km\,s^{-1}$ (our definition of a ``binary quasar'' from \S\ref{dat}),
using the estimator

\begin{equation} \label{wp}
\bar{W_{\rm p}}=\frac{QQ}{\langle QR \rangle}-1
\end{equation}

\noindent \citep[e.g.,][]{pe73,sh87,cs96}. Here, $QQ$ represents a count of
quasar-quasar data pairs and ${\langle QR \rangle}$ denotes the ``expected'' number of
quasar-random pairs in a given bin of redshift, angle or proper separation. Note that  

\begin{equation}
\langle QR \rangle = \frac{N_{\rm Q}}{N_{\rm R}}QR ~,
\label{eqn:QR}
\end{equation}

\noindent where $N_{\rm Q}$/$N_{\rm R}$ is the size of the quasar catalogue compared to the size of a (larger) random catalogue.
An appropriate random catalogue will mimic the angular and redshift distribution of the data, in the absence of any clustering. 
Since our KDE-complete sample of binary quasars is drawn from the KDE catalogue described in $\S$\ref{dat}, 
the random catalogue needs to have the same overall angular and redshift coverage as the KDE catalogue \citep[see, e.g.,][]{my06,my07}.

\begin{figure}
    \centering
       \includegraphics[angle=0,scale=0.3]{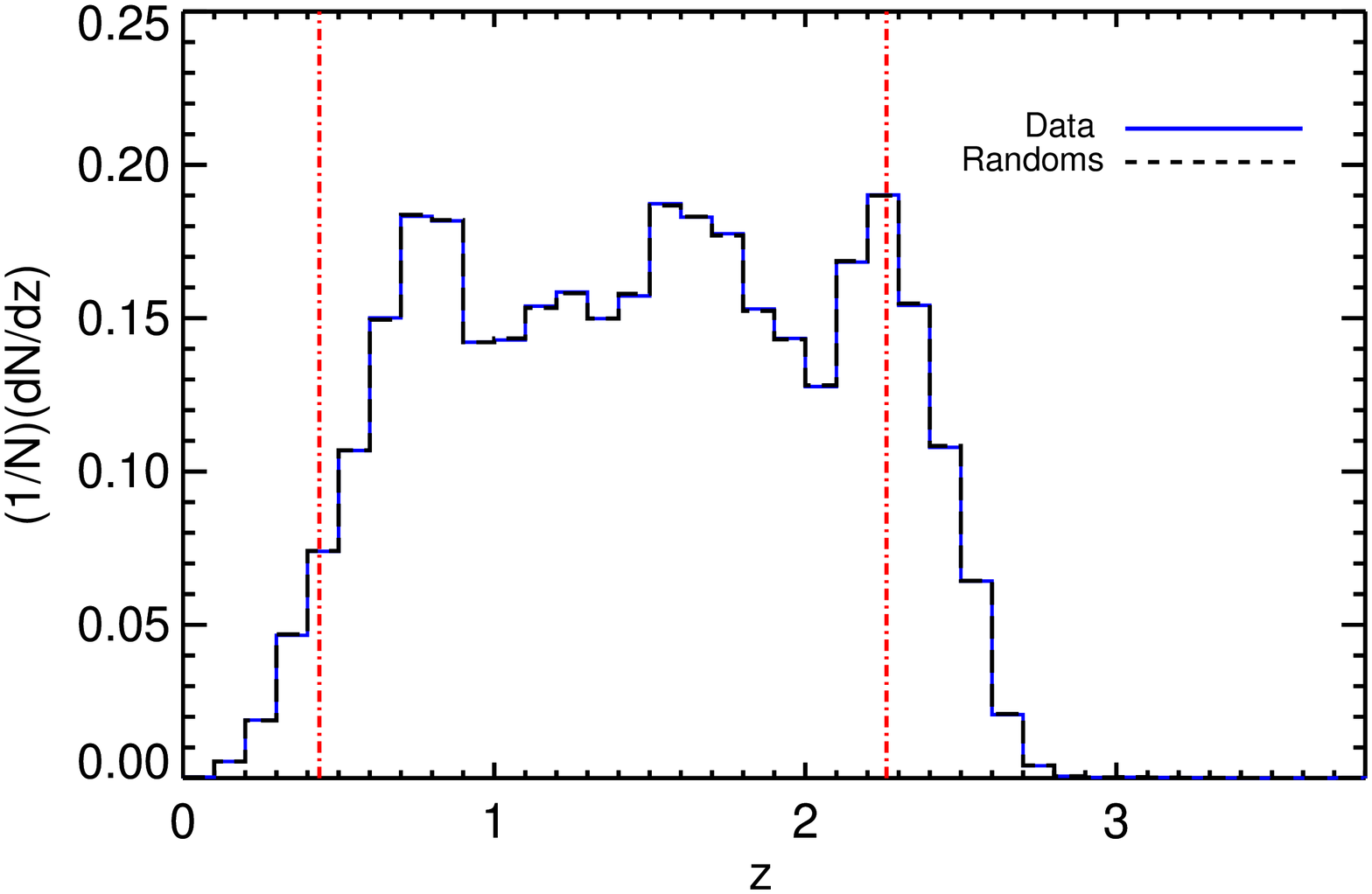}
    \caption{
The normalized redshift distribution of spectroscopically confirmed quasars in 
our sample of relevant pairs (blue solid line), compared to the generated 
distribution for our redshift random catalogue (black dashed line).
The vertical red dot-dashed lines delineate the redshift range of the 
KDE-complete sample. 
}\label{fig5:dist}
\end{figure}

\begin{table}
\centering
\begin{tabular}{cccccccc}

\hline
\hline
$\Delta z$   & & $(1/N)~dN/dz$  \\

\hline

0.43 &   &   0.191 \\
0.44 &   &   0.239 \\
0.45 &   &  0.200 \\
0.46 &   &   0.230 \\
0.47 &   &   0.240 \\
0.48 &   &   0.225 \\
0.49 &   &   0.239 \\
0.50 &   &   0.284 \\
0.51 &   &   0.288 \\
0.52 &   &   0.288 \\
0.53 &   &   0.305 \\
0.54 &   &   0.308 \\
0.55 &   &   0.258 \\
0.56 &   &   0.339 \\
0.57 &   &   0.303 \\
0.58 &   &   0.322 \\
0.59 &   &   0.389 \\
0.60 &   &   0.366 \\
0.61 &   &   0.398 \\
0.62 &   &   0.385 \\

\hline
\end{tabular}
\caption{Normalized distribution of the spectroscopic redshifts of quasars in our parent sample of candidates. The full table is available in the electronic version of this paper.}\label{tab:dndz}

\end{table}

The entire volume of the KDE catalogue comprises $\sim 41.93~ (h^{-1} \rm 
Gpc\,)^3$. Generating a sufficiently large random catalogue over such a volume 
purely for the purposes of making a kpc-scale clustering measurement is a computationally expensive task. Such an 
approach is also unnecessary, as we only seek $QR$ pairs with small angular 
separations ($\le 7.7\arcsec$). 
We therefore construct a random catalogue for our analysis using three independent steps.
As our sample of pairs is complete for proper scales of $17.0 \le R \le 36.2\,h^{-1}$\, kpc (see \S\ref{dat}), 
these three steps are sufficient to model the expected, unclustered distribution of our sample of binary quasars:

(1) We randomly selected a subset of $N_{\rm Q}=$ 342{,}581 KDE candidate quasars, corresponding to 
92.7\% of our parent sample of 369{,}559 KDE candidates (see \S\ref{dat}).
This down-sampling is necessary because 
the efficiency of the KDE algorithm for selecting our overall sample of candidate quasars ({\tt lowzts} and {\tt uvxts}; again see \S\ref{dat})
is $\sim$ 92.7\%. 

We randomly generated positions around these 342{,}581 KDE candidate quasars on angular scales of $2.9\arcsec < \theta < 7.7\arcsec$, which is the range of angular 
separations of candidate quasar pairs on our ``KDE-complete'' scales of interest (see \S\ref{dat} and, specifically Figure\,\ref{fig4:Rz}).
We will refer to the resulting catalogue as our {\em angular} random catalogue.

(2) Only $\sim$ 36\% (131{,}928) of the KDE candidates have a confirmed 
spectroscopic redshift. We used the {\it full} distribution of spectroscopic 
redshifts in the KDE sample, displayed in 
Figure \ref{fig5:dist}, and randomly drew redshifts
from the resulting ${\rm d}N/{\rm d}z$ for both those candidates with no 
spectroscopic redshift and for objects in our angular random catalogue.
Then, working with quasars with redshifts within our range of interest ($0.43 < z < 2.26$),
we down-sampled our angular random catalogue by retaining only random points in $QR$ pairs separated by 
$< 2000\,\rm km\,s^{-1}$ (our definition of a ``binary quasar'' from \S\ref{dat}).
We will refer to the resulting catalogue as our {\em redshift} random catalogue. 

\begin{figure}
\centering
\includegraphics[angle=0,scale=0.29]{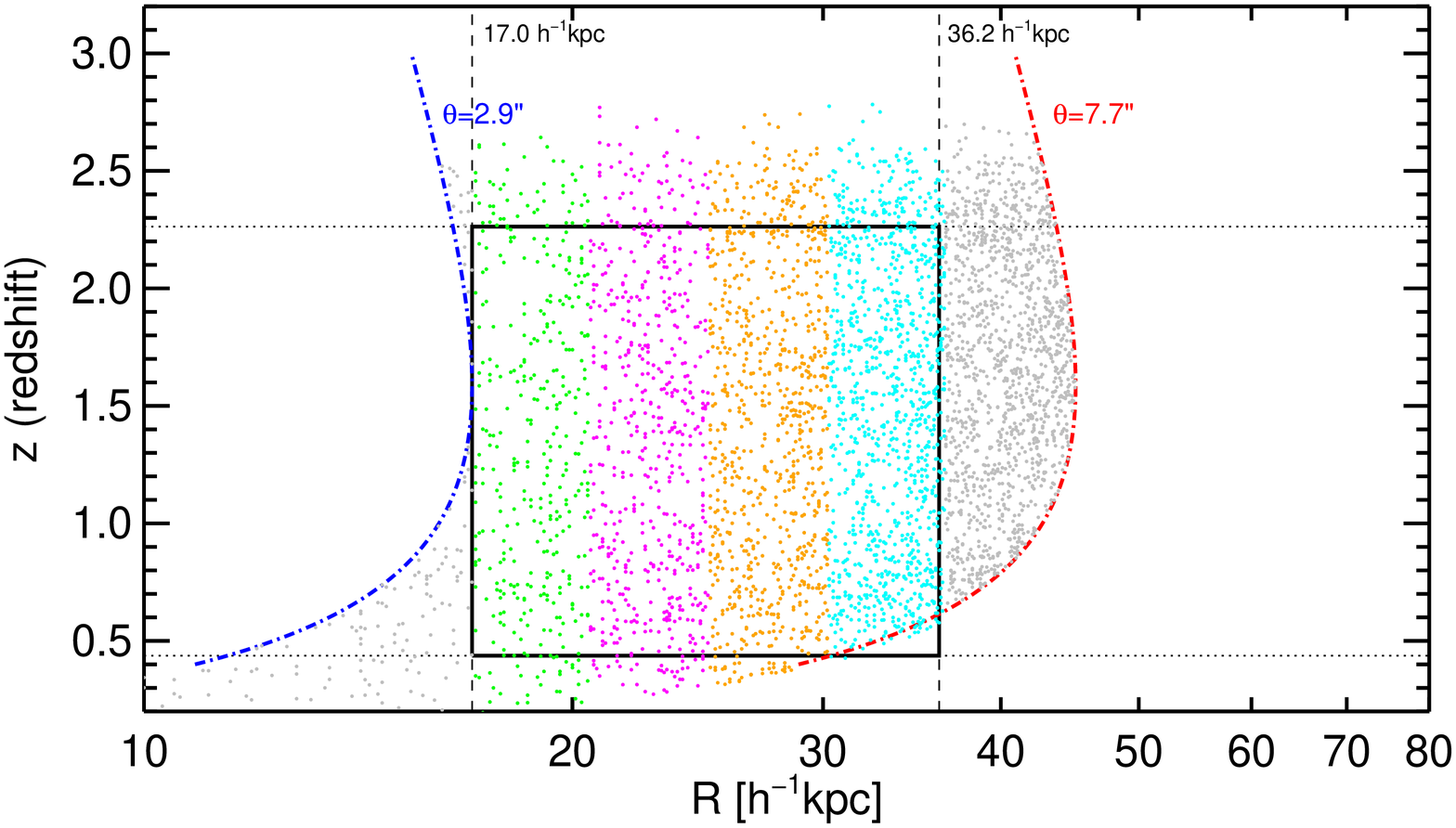}
\caption{The bins in proper scale that we use in our clustering measurement are shown in 
different colours. This figure illustrates the difference between how the random 
catalogue is populated in angle and the resulting random points that are counted 
in bins of proper scale.
The black box represents the limits of our KDE-complete sample (see also 
Figure\,\ref{fig4:Rz}).}\label{fig6:binran}
\end{figure}

(3) Using the redshift and angular separation information that we generated in steps (1) and (2), we further limited our redshift random catalogue to 
only $QR$ pairs that
intersected with the limits in redshift and proper scale of our ``KDE-complete'' sample of binary quasars (see Figure \ref{fig6:binran}). 
We will refer to the resulting catalogue as our {\em final} random catalogue. 
A total of 290{,}694 KDE candidate quasars have spectroscopic redshifts 
in the range of our complete sample of binary quasars ($0.43<z<2.2$).
The final random catalogue (``$R$'') can be used in conjunction with these
290{,}694 KDE candidate quasars (``$Q$'') to 
calculate $QR$ in Eqn.\,\ref{eqn:QR} as a function of scale or redshift.

The steps that produce the redshift and final random catalogues discard points that do not create eligible $QR$ pairs. It is therefore
necessary to generate a large enough initial angular random catalogue to retain a sufficiently large final random catalogue with which 
to infer $\langle QR \rangle$. 
We found that assigning each of the 290{,}694 KDE candidate quasars in our redshift range of interest
 $N={\rm 2000}$ random points on scales of $0\arcsec < \theta < 7.7\arcsec$\footnote{We
assigned 2000 points over $\theta < 7.7\arcsec$ and then clipped them to cover $2.9\arcsec < \theta < 7.7\arcsec$, to provide 
flexibility if our minimum angle changed.} was sufficient in this regard, 
as such a schema ultimately provided more than 20 random points around each KDE candidate quasar. Essentially, this means that our {\em final} random
catalogue is at least 20$\times$ larger than our data catalogue.

An important consideration is that the $N_{\rm R}$ in Eqn.\,\ref{eqn:QR} does not denote 
the $N={\rm 2000}$ random points that we generated around each of the $N_{\rm Q}={\rm 290{,}694}$ KDE candidate quasars in our redshift range of interest.
Rather, it corresponds to the number of random points that would have truly been generated, 
had we chosen to populate the {\em entire survey volume}. We calculate $N_{\rm R}$ as  the {\it``populated 
areal number density of the random points''} $\times$ {\it``the full area of the survey footprint''}:

\begin{equation}
N_R = \frac{N}{A(< 7.7\arcsec)}A_{\rm full} ~,
\label{eqn:Afull}
\end{equation}

\noindent where $N={\rm 2000}$ is the number of random  points we generated around each candidate quasar to $\theta < 7.7\arcsec$, 
$A(< 7.7\arcsec)$ is the survey area within 7.7\arcsec of a candidate quasar and $A_{\rm full}$ is the full area of the survey footprint  (the orange footprint in 
Figure \ref{fig3:aitoff}).

To calculate the survey area, we use the SDSS ``survey coordinates'', $\eta$ and $\lambda$ \citep[e.g.][]{Sto02}, to construct 
stripe-shaped polygons along great circles using the {\sc mangle} software \citep{bl03,teg04,sw08}. We also create ``holes" in the footprint corresponding to 
SDSS imaging masks\footnote{e.g., http://classic.sdss.org/dr6/products/images/index.html}. Note that when we created the angular random catalogue, we
discarded any points that lay in holes or outside of the survey area, but this made very little difference on scales of $\theta < 7.7\arcsec$.
Based on this process, the total area of the survey footprint that is used in this study is $A_{\rm full} = 7600.4\,{\rm deg^2}$. Since we only consider angular scales up to $7.7\arcsec$ the ``effective'' area around any individual candidate quasar is $A(< 7.7\arcsec) = 1.44 \times 10^{-5}\,{\rm deg^2}$. So, $N_{\rm R} = (2000 \times 7600.4)/1.44 \times 10^{-5} \sim 10^{12}$. In other words, the process that we have outlined would be equivalent to generating a very, very large random catalogue across the entire survey volume.

\subsection{Theoretical Considerations}\label{W2w}
The volume averaged projected correlation function ($\bar W_{\rm p}$) is a useful estimator for our purposes given the large volume occupied by quasars over a wide redshift range,
compared to the small scales on which we seek to measure clustering. $\bar W_{\rm p}$ can
be converted to the more common clustering estimators used on large scales
via the formalism presented in, e.g., \citet{he06a}.

The projected real-space correlation function of quasars with a maximum velocity difference of $|\Delta v|< 2000\rm km\,s^{-1}$ can be interpreted as :
\begin{equation}\label{wp}
w_{\rm p}(R,z)=\int_{-v_{\rm max}/H(z)}^{v_{\rm max}/H(z)} \xi_{s}(R, s, z)\,{\rm d}s,
\end{equation} 
where $v_{\rm max}=2000\rm{\,km\,s^{-1}}$, $H(z)$ is the expansion rate at redshift $z$ and $\xi_{s}$ is the quasar correlation function in redshift-space.

As discussed in \citet{he06a}, it is a good 
approximation to replace the redshift-space correlation function $\xi_{s}$ with 
its three-dimensional real-space counterpart $\xi(r)$.
We measure the volume-averaged correlation 
function $\bar W_{\rm p}(R_{\rm min}, R_{\rm max}, z)$ (abbreviated to $\bar 
W_{\rm p}(z)$), by integrating over the entire radial bin of proper 
distance $[R_{\rm min}, R_{\rm max}]$

\begin{equation}\label{wpbar}
\bar W_{\rm p}(z)= \frac{\int_{-v_{\rm max}/H(z)}^{v_{\rm max}/H(z)} \int_{R_{\rm min}}^{R_{\rm max}} \xi(R, x, z) ~ 2\pi R\,{\rm d}R\,{\rm d}s}{V_{\rm shell}}
\end{equation} 

\noindent where $\xi(R, x, z)$ is the correlation function
and $V_{\rm shell}$ is the volume of the cylindrical shell in redshift space over which we integrate

\begin{equation}\label{vol}
V_{\rm shell}=\pi (R_{\rm max}^{2}-R_{\rm min}^{2}) \left[\frac{2v_{\rm max}}{H(z)}\right] , 
\end{equation} 

\noindent and then averaging the redshift-dependent $\bar W_{\rm p}$ in Eqn.\,\ref{wpbar} over the 
redshift distribution of quasars in our sample. 

We need to average $W_{\rm p}(z)$ over the redshift distribution of our sample in a given redshift bin, in order to compare to our clustering measurement.
To estimate the redshift distribution for
our quasars of interest in any slice of redshift, we use the Pure Luminosity Evolution (PLE) model of \citet{cr09}
with $\alpha = -3.33 ,~ \beta = -1.42,~ \rm M^{\star} = -22.17$ and $\rm
log(\rm \phi^{\star}) = -5.84 \, \rm Mpc^{-3}~mag^{-1}$. We adopt this particular luminosity function as the sample of quasars studied in \citet{cr09} is a reasonable
match ($0.4 < z < 2.6$) to the redshift range of our sample, and extends well beyond ($g < 21.85$) our magnitude limit.

Because we measure $\bar W_{\rm p}$ for {\em quasars} the $\xi$ included in Eqn.\ \ref{wpbar} is
typically the correlation function of quasars, which we will denote $\xi_{\rm Q}$. We will adopt two typical theoretical forms for  this
function. First, a two-parameter power-law of the form 

\begin{equation}\label{eqn:powerlaw}
\xi_{\rm Q}(r)=(r/r_{0})^{-\gamma}
\end{equation}

\noindent where $r_0$ is the {\em correlation length}, defined as the most common (probable) separation 
between two quasars in the sample, and $\gamma$ is the exponent that best recreates
the shape of quasar clustering. Second, 

\begin{equation}\label{eqn:Apar}
\xi_{\rm Q}(r)=A ~\xi_{\rm }(r)
\end{equation}


\noindent where $A$ is the ratio of the clustering amplitude of quasars to that of the underlying dark matter distribution and
 $\xi_{\rm }(r)$ is the correlation function of underlying dark matter, for which we adopt the
 model of \citet{sm03}. In some places in \S\ref{dis}, we use values of 
 $r_{0}$ and $\gamma$, or a form for $\xi(r)$, that have been derived for the clustering
 of quasars or dark matter on Mpc-scales. We then use Eqn.\,\ref{wpbar} to project this Mpc-scale
 result down to our kpc-scales of interest.
 
Phenomenologically, the formalism of Eqn.\,\ref{eqn:Apar} resembles that for the bias of tracers of dark matter \citep[e.g.,][]{kai84}. We appreciate, though, 
that small-scale bias could change rapidly with scale, and that the amplitude of quasar clustering is likely to be a complex function of several factors on non-linear scales.
Any association we make between the parameter $A$ and the bias of dark matter ($b_{\rm Q}$) in this work, therefore, is only for to make comparisons
between the amplitude of quasar clustering at kpc- and Mpc-scales. In essence, we adopt Eqn.\ \ref{eqn:Apar} only as an empirical 
parameterization of the amplitude of quasar clustering on kpc-scales. We reserve models that have a more complex physical interpretation for a later paper.
 
\section{Results and discussion}\label{dis}

Our KDE-complete sample of confirmed binary quasars is $\sim 6$ times larger than 
any individual previous sample, allowing 
us to measure the scale-dependence of quasar clustering 
at $\lesssim 40 ~ h^{-1}$\,kpc with unparalleled precision. In addition, our large
sample extends across multiple bins in redshift that each contain about as many binary quasars as
any previous sample. This allows us to study the evolution of quasar clustering on these very small 
scales for the first time.

\subsection{The scale-dependence of $\bar W_{\rm p}$ at $\lesssim 40 ~ h^{-1}$\, kpc}\label{rdep}

\begin{figure*}
\includegraphics[angle=0,scale=0.35]{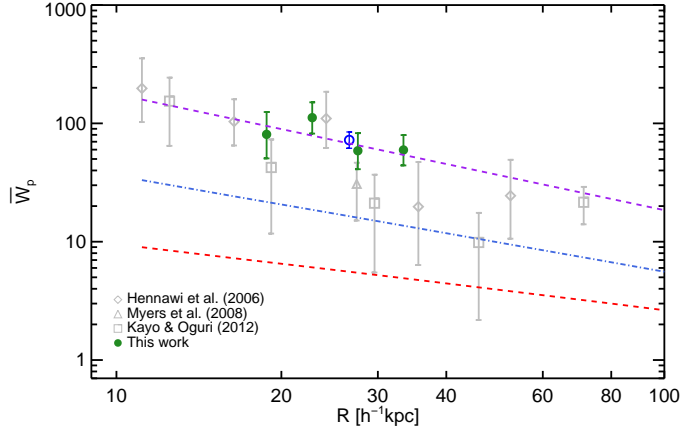}
\caption{The projected correlation function of quasars in four bins of proper 
scale (filled green circles) as well as the measured correlation function for 
the full sample at $\bar{z}=1.55$ and $R=26.6\,h^{-1}$\,kpc (open blue circle). 
The grey symbols depict similar measurements from previous studies. Where necessary, we have converted the comoving coordinates used in previous measurements to proper coordinates 
in order to compare with our data. The red and 
blue dashed and dot-dashed lines are the extrapolation of the $\bar W_{\rm p}$ 
correlation function reported on Mpc-scales by 
\citet{po04} with ($\gamma=1.53$, $r_0=4.8 ~h^{-1}$\,cMpc) and ($\gamma=1.8$, 
$r_0=5.4 ~h^{-1}$\,cMpc), respectively. 
The purple line shows the best fit to the green data points assuming a 
correlation length of $r_0=5.0\,h^{-1}$\,cMpc. This line has a power-law index of $\gamma=1.97\pm0.03$ indicating that the correlation function is steeper on kpc-scales than has been estimated for many quasar samples on Mpc-scales (as discussed further in \S4.1)} \label{fig7:wpr}
 \end{figure*}

We measure the volume-averaged projected correlation function ($\bar W_{\rm p}$) 
of quasars in four bins of proper scale centered at 18.8, 22.8, 27.6 and 
33.4\,$h^{-1}$\,kpc which 
contain 7, 14, 11 and 15 binary quasars, respectively. 
The bins were chosen to have the same width in logarithmic scale. 
The measured $\bar W_{\rm p}$ for each bin of proper separation together with the measured
$\bar W_{\rm p}$ for the full sample at $\bar{z}=1.55$ and ${\bar R}=26.6\,h^{-1}$\,kpc are shown in Figure\,\ref{fig7:wpr}. Multiple past works have argued that pair counts on small scales are independent, and that clustering on these scales can be adequately described by a Poisson distribution \citep[e.g.][]{sb94,cs96,my06,ch12,ch13}. We therefore adopt Poisson errors from \citet{geh86} for our measurements of $\bar W_{\rm p}$.

Table\,\ref{tab:wpval} lists our measured $\bar W_{\rm p}$ in each bin of proper separation and
for our full, KDE-complete sample of 47 binary quasars.

\begin{table}
\centering
\begin{tabular}{cccccccc}
\hline
\hline
$R$ ($h^{-1}$\,kpc)& & & & & & & $\bar W_{\rm p}$ \\
\hline 
\vspace{2pt}
18.8 &  & & & &  & & $79.8^{+43.5}_{-29.8}$\\ 
\vspace{2pt}
22.8 &  & & & &  & & $109.1^{+38.0}_{-29.1}$\\  
\vspace{2pt}
27.6 &  & & & &  & & $58.0^{+23.7}_{-17.5}$ \\ 
33.4 &  & & & &  & & $59.2^{+19.9}_{-15.4}$\\
\hline

26.6 &  & & & &  & & $72.28^{+15.2}_{-13.5}$\\
\hline
\end{tabular}
\caption{The volume-averaged correlation function for the four bins of proper 
separation displayed in Figure\,\ref{fig7:wpr}. The last row corresponds
to the full range of scales (the open blue circle in Figure\,\ref{fig7:wpr}).}\label{tab:wpval}
\end{table}

Figure\,\ref{fig7:wpr} also compares our measurement of $\bar W_{\rm p}$ to  
previous estimates of quasar clustering on small scales at redshifts of $0.5 \lesssim z \lesssim 2.5$.
\citet{he06a} constructed a large, homogeneous catalogue of binary quasars from SDSS DR3 and used a sub-sample of them to
measure quasar clustering on small scales. The clustering sub-sample of \citet{he06a}
included 23 binary quasars on proper scales of $10 \lesssim R \lesssim 100\,h^{-1}$\,kpc.
\citet{my08} built on this work by discovering 10 new binary quasars in the SDSS DR4 KDE catalogue, and used them to study quasar clustering on 
a specific range of very small proper scales ($20 \lesssim R \lesssim 30\,h^{-1}$\,kpc).
More recently, \citet{ko12} compiled a sample of binary quasars from observations 
conducted across SDSS DR7 as part of the SDSS Quasar Lens Search \citep[e.g.][]{in12}, and used
26 binaries with comoving separations of 10 -- 200\,$h^{-1}$\, ckpc  
(proper scales of $5 \lesssim R \lesssim 100\,h^{-1}$\,kpc) to measure $\bar W_{\rm p}$.
The sample of \citet{ko12} only shares 4 binaries with that of \citet{he06a} and a further 2 with \citet{my08}. This is largely because the sample of \citet{ko12}
is more complete than the sample of \citet{he06a}, covers a larger range of scales than the sample of \citet{my08}, and covers a larger
portion of the SDSS footprint as compared to both studies.

The sample of \citet{ko12} contains only 4 binary quasars on scales
of $17 \lesssim R \lesssim 36\,h^{-1}$\,kpc, and two of these are the pairs that \citet{ko12} incorporated from 
\citet{my08}. Further, the clustering sub-sample of \citet{he06a}
includes only 8 binary quasars on proper scales of $17 \lesssim R \lesssim 36\,h^{-1}$\,kpc.
Our KDE-complete sample of binary quasars is thus $\sim6\times$ larger than any previous
statistically homogeneous sample at $R \sim 25\,h^{-1}$\,kpc
and so can substantially improve the accuracy
of quasar clustering measurements on small scales.
The blue open circle in Figure\, \ref{fig7:wpr} shows the statistical significance of 
our measurement compared to recent such measurements on kpc-scales. 
Our KDE-complete sample is about 4$\times$ larger than all other combined 
samples at $17 \lesssim R \lesssim 36\,h^{-1}$\,kpc. This essentially means that our results
can be used to improve constraints on kpc-scale
quasar clustering by a factor of 2 compared to previous work. 

The real-space correlation function of quasars can be modeled by a simple power-law (see Eqn.\,\ref{eqn:powerlaw}).
Quasars in our redshift range of interest ($0.4 < z < 2.3$) have been argued to have a 
range of power-law indexes based on clustering measurements conducted on Mpc-scales. The sample that best
matches our luminosity and redshift range ($b_{\rm J} < 20.85; 0.3<z<2.2$) is that of the
2dF QSO Redshift Survey \citep[2QZ;][]{cr04}. For the 2QZ,  
\citet{po04} measured a best-fit power-law of $\gamma=1.53$ for $r_0=4.8\,h^{-1}$ cMpc, 
rising to $\gamma=1.8$ for $r_0=5.4 ~h^{-1}$\, cMpc. \citet{ro09} found that 
the clustering of brighter quasars from the
SDSS ($i < 19.1; 0.3 < z < 2.2$) required a steeper power-law of 
$\gamma=1.90^{+0.04}_{-0.03}$ for $r_0=5.5\,h^{-1}$ cMpc. The red dashed and blue dot-dashed lines
in Figure\,\ref{fig7:wpr} compare the best-fit power laws from \citet{po04} to our
results, and it is clear that our data necessitate a much steeper power-law.
Fixing the correlation length to $r_0=5\,h^{-1}$ cMpc, we use a maximum likelihood fitting
procedure to determine that our data require a
power-law index of $\gamma = 1.97\pm0.03$, which is plotted as the purple dashed line in Figure\,\ref{fig7:wpr}.
This power-law index is far in excess of the results of \citet{po04} but is in reasonable
agreement with the Mpc-scale clustering of substantially brighter SDSS quasars from \citet{ro09}.
Our results also support the study of 
\citet{ko12}, who found a power-law of $\gamma=1.92\pm0.04$ for $r_0=5.4\,h^{-1}$\, cMpc  from
a study of the clustering of bright ($i < 19.1; 0.6 < z < 2.2$) SDSS quasars on proper scales
of $4 \lesssim R \lesssim 85\,h^{-1}$\,kpc.

At low redshift ($z\sim0.5$ and below), quasars appear to be roughly unbiased \citep[e.g.][]{cr05},
and cluster similarly to $L^*$ galaxies. Given that quasars are thought to be
merger-driven \citep[e.g.][]{hop06,Hop07}, it is interesting
to compare the overall shape of the correlation function of galaxies and of quasars at similar
redshift. Any excess in quasar clustering compared to galaxies might indicate that
quasars ignite in particularly grouped or ``merger-prone'' environments \citep[again see][]{Hop07}.
For example, \citet{wat10} suggest that enhanced quasar activity by mergers 
might be responsible for the shape differences between the correlation function 
of Luminous Red Galaxies (LRGs) and quasars on very small scales.
Large spectroscopic galaxy surveys are now approaching $z\sim1$, so it is becoming realistic to
compare quasar and galaxy correlation functions in similar redshift ranges.

Recent galaxy clustering results on Mpc-scales, tend to find power-law slopes that are shallower
than $\gamma = 2$. For example, \citet{Fav16} find $\gamma=1.6\pm0.1$ in redshift-space
for $s_0=(5.3\pm0.2)\,h^{-1}$\, cMpc  for Emission Line Galaxies at $z\sim0.8$. \citet{Coi16}
find a range of power-law slopes for blue and red galaxies from the PRIMUS survey over the redshift range
$0.4 \lesssim z \lesssim 0.9$. For populations that have $r_0$ consistent (within their
1-$\sigma$ errors) with $5\,h^{-1}$\, cMpc  \citet{Coi16} find $\gamma=1.6$ -- 1.7, with an error less than $\pm0.1$.
On smaller scales, galaxy clustering may steepen, however.
\citet{ma06} tracked the clustering of
$z\sim0.25$ SDSS LRGs down to scales of $\sim 10\,h^{-1}\,\rm kpc$ and
estimated $\gamma\sim2.0$, although they also found a large correlation length of $r_0\sim10\,h^{-1}$\,cMpc. 
\citet{ze11} found power-law slopes ranging from $\gamma \sim1.8$ -- $2.0$ for $r_0\sim4.5$ -- $10.4\,h^{-1}$\, cMpc  down 
to comoving distances of $\sim 100 ~h^{-1}\,\rm ckpc$ for SDSS galaxies at 
$z \lesssim 0.25$. For samples that have $r_0$ in the range $4.5$ -- $5.5\,h^{-1}$\, cMpc  \citet{ze11} find
$\gamma \sim 1.8$ -- 1.9. More recently and at higher redshift, \citet{Zha16} find $\gamma \sim1.95$
down to scales of $\sim 300\,h^{-1}\,\rm ckpc$
for the clustering of $z\sim0.7$ LRGs drawn from the SDSS-IV/extended Baryon Oscillation Spectroscopic Survey (eBOSS).
Our inferred power-law of $\gamma = 1.97\pm0.03$ for $r_0=5\,h^{-1}$\, cMpc  is therefore at the steeper
end of what has been measured for galaxies, but is not inconsistent with measurements at higher redshift that
sample smaller scales. A detailed theoretical analysis, 
such as the Halo Occupation Distribution \citep[HOD;][]{bw02} formalism, should be 
able to use our measurements to better quantify whether quasar clustering  
exceeds galaxy clustering on kpc-scales, or whether quasars occupy similar halos to 
certain types of galaxies. We defer such a detailed HOD analysis to a later paper.

\begin{figure*}
    \centering
    \begin{subfigure}{
        \centering
        \includegraphics[angle=0,scale=0.3]{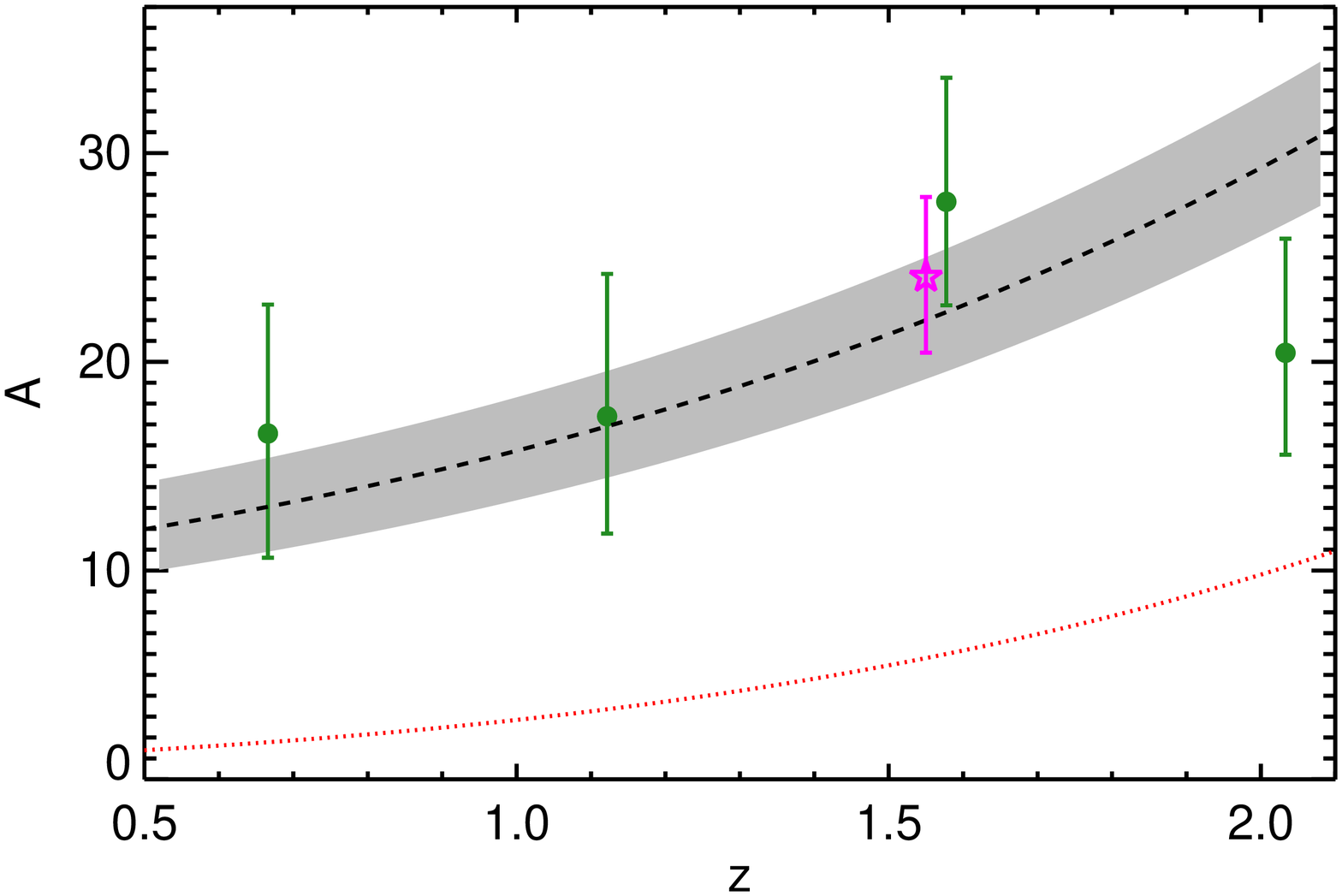}
        \includegraphics[angle=0,scale=0.3]{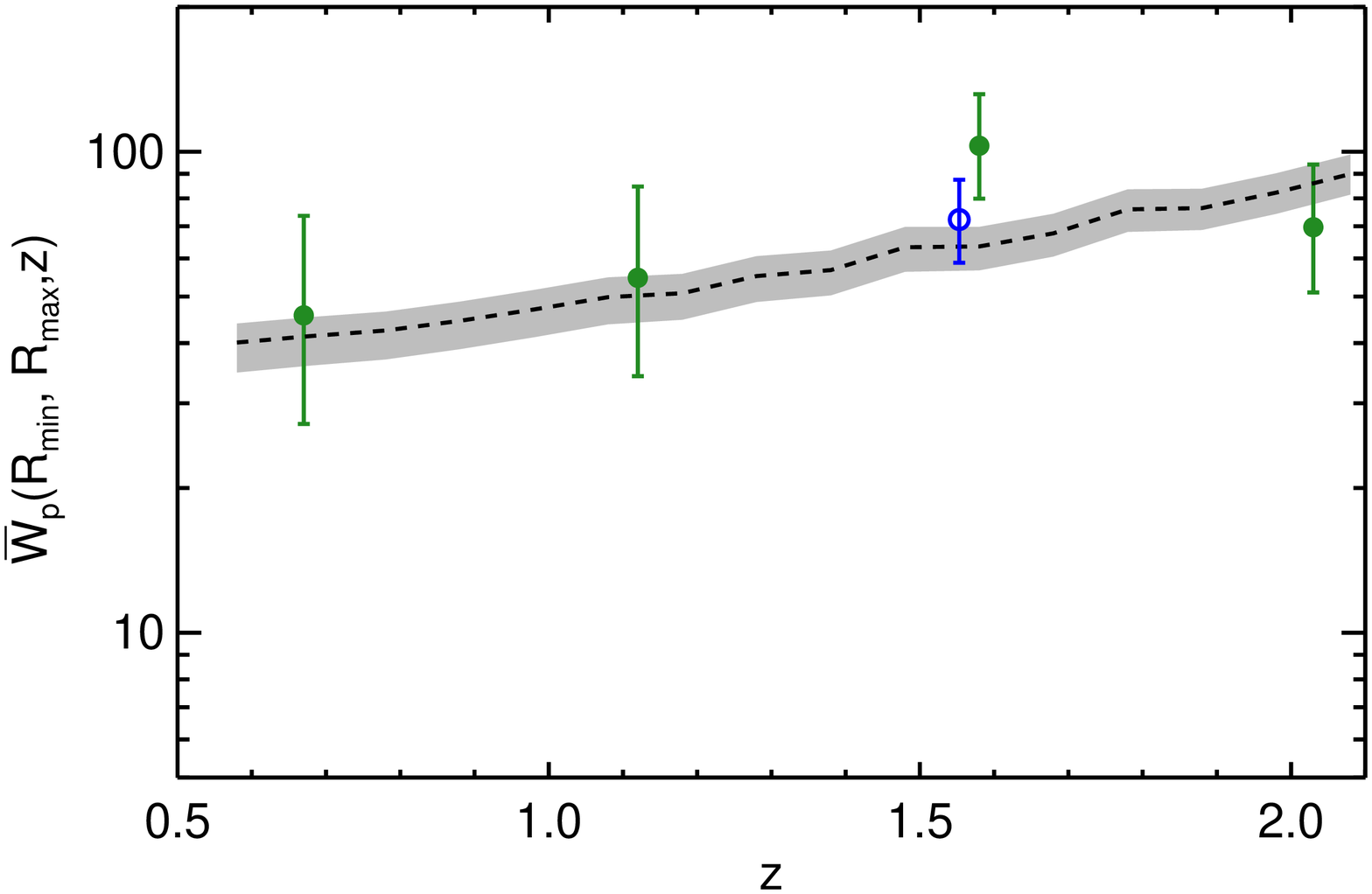}}
        \caption{Left: Our deduced quasar clustering amplitude at $R\sim25\,h^{-1}$\,kpc in each bin of 
redshift. The pink star depicts the amplitude derived from our measurement of quasar clustering for our
full sample $\left(A=24.1\pm3.6\right)$. The red dotted line is the model 
for the evolution of quasar bias on Mpc-scales proposed by \citet{cr05}; $b_{\rm Q}^{2}(z)=(0.53+0.289 (1+z)^2)^2$.
The black dashed line depicts the best-fit value we find for a 
one-parameter fit of $A(z)=(c+0.289 (1+z)^2)^{2}$, which is
$c=2.81\pm0.31 $. The grey envelope depicts the 1-$\sigma$ confidence interval for the fitted parameter $c$. Right: The projected correlation function of quasars 
in four bins of redshift. Each redshift bin spans the full range of proper scales of our KDE-complete sample ($17.0\le R \le 
36.2\,h^{-1}$\,kpc). The black dashed line is the calculated $\bar W_{\rm p}$ for the full redshift range of our KDE-complete sample, using the model from the left-hand panel. 
The grey envelope is the translation of the confidence intervals from the left-hand panel
based on the relationship between $A$ and $\bar W_{\rm p}$ outlined in \S\ref{W2w}.}\label{fig8:bzwz}
    \end{subfigure}
\end{figure*}

\subsection{Redshift dependence of $\bar W_{\rm p}$}\label{zdep}

Measurements of the evolution of quasar clustering on Mpc scales \citep[e.g.][]{cr05}, in combination with
the quasar luminosity function, have 
helped to constrain fueling models for quasars and provided a framework to link quasar activity 
to galaxy formation \citep[see, e.g., ][and references therein]{Hop07}. 
Broadly, the quasar correlation length on Mpc-scales
does not appear to evolve by more than a factor of $\sim 2$ over the 
range $0.5 \lesssim  z \lesssim 2.5 $ \citep[see, e.g.,][and references 
therein]{ef15}. This, in turn, implies that quasar bias increases significantly between
redshift 0.5 and 2.5, and that the characteristic mass of the dark matter haloes that host
quasars is roughly constant across this redshift range.
\citet{my07b} estimated how quasar clustering on small scales changes with redshift using 
a sample of 91 photometrically classified candidate quasars and
found that UV-excess quasars at $28~ h^{-1}$\,kpc cluster $>5$ 
times ($\sim 2 \sigma$) higher at $z > 2$ than at $z < 2$. However, 
the evolution of quasar clustering on proper scales of $< 50\,h^{-1}$\,
kpc has not yet been measured using a {\em spectroscopically confirmed} sample of quasar pairs, likely because
sample sizes have never been sufficiently large to bin by redshift. With the
unprecedentedly large number of binary quasars in, 
and wide redshift range 
of, our KDE-complete sample, we can make this measurement for the first time.

We divide our KDE-complete sample of quasar pairs into 
four bins of redshift of similar width ($\Delta z \simeq 0.46$) centered at $z = 0.67, 
1.12, 1.58$ and $2.03$. These bins contain 6, 7, 20 and 14 quasar pairs,  
respectively\footnote{Choosing the redshift slices such that they contain the same number of pairs
would cause some bins to be very narrow.}. We then measure the correlation function $\bar W_{\rm 
p}(R_{\rm min},R_{\rm max})$ in each bin of redshift over the full range of proper scales of  
our sample ($17.0 < R < 36.2 ~h^{-1}$\,kpc). We plot the
results of this analysis in the right-hand panel of Figure \ref{fig8:bzwz}. 
Having measured the volume-averaged correlation function in four slices of 
redshift, we use the method described in $\S $\ref{W2w} to derive the amplitude of quasar clustering (A from Eqn.\,\ref{eqn:Apar}) in each bin of redshift.
The left-hand panel of Figure \ref{fig8:bzwz} shows the
 values of A that correspond to the measured $\bar W_{\rm p}(z)$ values plotted in the right-hand panel.  
 We measure the clustering amplitude of quasars at $\sim 25~ h^{-1}$\,kpc from our full KDE-complete sample of 47 confirmed 
binaries to be $A =24.1\pm3.6$ (the pink star in Figure\,\ref{fig8:bzwz}).  

\citet{cr05} measured a clustering amplitude equivalent to $A\sim5$ at $z\sim1.5$ on Mpc-scales. 
The fact that we find a factor of $\sim4\times$ stronger amplitude for quasar clustering on kpc scales than has been
found on Mpc scales suggests that, on small scales, quasar clustering climbs rapidly above
the dark matter model \citep{sm03} that we use in Eqn.\ \ref{wpbar}. This was interpreted as an ``excess'' by \citet{he06a} and
\citet{my08}, perhaps driven by pairs of quasars being fed during galaxy mergers. \citet{Hop07} argued instead that
strong quasar clustering on small scales is simply
indicative of quasars occupying group-scale or ``merger-prone'' environments. More recently, the
small-scale clustering of quasars has been modeled using the ``one-halo" term in
the HOD \citep[e.g.,][]{ko12,ric12,ric13}. As we argue in \S\ref{rdep}, this ``excess'' is, in fact, probably close-to-consistent with
the amplitude of clustering found for some types of galaxies on small scales. 

Our unprecedentedly precise measurements of $\bar W_{\rm p}$ 
on scales of $\sim 25~ h^{-1}$\,kpc allow us to make a first comparison of 
the {\em evolution} of quasar clustering over 3 orders of magnitude in scale.
To do so, we compare our measurements to the 
empirical description of the evolution of quasar clustering derived by \citet{cr05}
over scales of $1 < s <25 ~h^{-1}\,\rm cMpc$. Using our empirical formalism from Eqn.\,\ref{eqn:Apar}, \citet{cr05}
found the equivalent of $A(z)=\left[0.53+0.289 (1+z)^2\right]^2$.
Our goal is to compare the evolution of the amplitude of quasar clustering on kpc- and Mpc-scales.
Because we measure a larger amplitude ($A$) on kpc-scales
than is found on Mpc-scales, we allow the offset in the \cite{cr05} empirical
description to float and fit a model of the form $A=\left[c+0.289 (1+z)^2\right]^2$.
We find a best fit of $c=2.81\pm 0.31$ to our  measurements in four slices of redshift over the range $0.43<z<2.26$, 
which we plot in (both panels of) Figure\,\ref{fig8:bzwz}. 
We find that the evolution of the amplitude of quasar clustering on kpc-scales across a wide range of redshift, 
is in reasonable agreement with the overall Mpc-scale empirical description of
\citet{cr05}, once we account for the amplitude offset of a factor of $\sim 4\times$. 
The $\chi^2$ value of our best fit is 4.2, which is only rejected at a confidence-level of
12\%. Based on our, admittedly highly empirical model of Eqn.\,\ref{eqn:Apar}, this suggests that the evolution
of the amplitude of quasar clustering on the smallest scales can be adequately modeled using descriptions of quasar evolution on Mpc-scales.

\section{Summary and Conclusions}\label{sum}

We present by far the largest sample of spectroscopically confirmed 
binary quasars with proper transverse separations of $17.0\le R \le 36.2\,h^{-1}$\,kpc. 
Our sample, which is $\sim6\times$ larger than any previous homogeneously selected sample
on these proper scales, is 
derived from SDSS imaging over an area corresponding 
to SDSS DR6. Our quasars are targeted using a Kernel 
Density Estimation technique (KDE), and confirmed using long-slit spectroscopy 
on a range of facilities. We derive a statistically complete sub-sample of 47 binary quasars with 
$g<20.85$, which extends across angular scales of $2.9\arcsec < \Delta \theta < 6.3\arcsec$ and
redshifts of $0.43<z<2.26$. This sample is targeted from 
a parent catalogue that would be equivalent to a full spectroscopic survey of 
nearly 360,000 quasars.

We determine the projected correlation function ($\bar W_{\rm p}$) of $0.43<z<2.26$
quasars over proper transverse scales of $17.0\le R \le 36.2\,h^{-1}$\,kpc, 
in four bins of scale. We find that quasars
cluster on kpc-scales far higher than implied by a $\gamma=1.8$ power-law, as has been
adopted by some authors on Mpc-scales \citep[e.g][]{po04}. For $r_0=5\,h^{-1}$\,cMpc, we
find that a power-law slope of $\gamma = 1.97\pm0.03$ is therefore required to fit quasar clustering on
proper scales of $R \sim 25\,h^{-1}$\,kpc. This is steeper than what is typically measured for galaxies,
but is consistent with some measurements of galaxy clustering, particularly on very small scales and
at $z > 0.5$. We therefore confirm previous results that suggest that the steep shape of quasar clustering on small scales may
be indicative of quasars ``turning on'' in galaxy mergers \citep[e.g.][]{he06a,my08} or of quasars
inhabiting group-scale (``merger-prone'') environments \citep[e.g.][]{Hop07}. The $\gamma\sim2$ power-law
we find is also consistent with results that suggest that quasars require a steeper 
power-law index than is typical for popular
theoretical dark matter density relations \citep[e.g.][]{mo96,nfw97}. A full modeling of this effect will
require an in-depth study of the ``one-halo" term of
the HOD \citep[e.g.,][]{ko12,ric12,ric13}, which we reserve for future work.

Our sample of binary quasars is the first that is sufficiently large to study quasar clustering as a function of redshift 
on proper scales of $R \sim 25\,h^{-1}$\,kpc.
To investigate the evolution of quasar clustering on small scales, we measure the projected quasar correlation function in four 
bins of redshift over $0.4 \le z \le 2.3$ and derive the amplitude of quasar clustering on small scales. We find that, at $z\sim1.5$, the clustering of quasars
substantially exceeds our chosen dark matter model \citep{sm03}, and does so by a factor of about 4 in amplitude as
compared to the excess-over-dark-matter on Mpc-scales.

We compare the evolution of the amplitude of quasar clustering on proper scales of $R \sim 25\,h^{-1}$\,kpc to empirical relationships
derived by \citet{cr05} on scales of $\sim 10\,h^{-1}$\,Mpc. 
Our kpc-scale results cannot rule out descriptions of the evolution of
quasar clustering on Mpc-scales, which, at its simplest would imply that the dark 
matter in which quasars are embedded evolves similarly to the baryonic matter over 3 orders of magnitude in scale.
However, our sample size is too small and our modeling is too physically simplistic to formally detect a strong evolution in quasar bias from $z\sim 0.5$ to $z\sim2.5$,
which leaves open the possibility that how the clustering of quasars evolves with redshift may be a function of scale.

\section*{Acknowledgments}
SE and ADM were partially supported by the National Science Foundation (NSF) through 
grant number 1515404. SGD, AAM, and MJG acknowledge partial support from NSF grants AST-1313422, AST-1413600, and AST-1518308.  We thank the staff of Palomar Observatory for their assistance during our observing runs. Observations reported here were obtained at;
(1) the MMT Observatory, a joint facility of the Smithsonian Institution and the University of Arizona;
(2) the Hale Telescope, Palomar Observatory as part of a continuing collaboration 
between the California Institute of Technology, NASA/JPL, Oxford University, 
Yale University, and the National Astronomical Observatories of China; and
(3) the Mayall telescope at Kitt Peak National Observatory, National Optical 
Astronomy Observatory (NOAO Prop. ID: 2008A-0127; PI: Myers), which is operated by the 
Association of Universities for Research in Astronomy (AURA) under cooperative 
agreement with the National Science Foundation. The authors are honored to be 
permitted to conduct astronomical research on Iolkam Du'ag (Kitt Peak), a 
mountain with particular significance to the Tohono O'odham.
This work used the facilities of The Advanced Research Computing Center at the University
of Wyoming (Advanced Research Computing Center. 2012. Mount Moran: IBM System X cluster. Laramie, WY: University of Wyoming. 
{\small \url{http://n2t.net/ark:/85786/m4159c}}).

\bibliographystyle{mn2e.bst}
\bibliography{referencelist}

\onecolumn
\onecolumn
\footnotesize 
 \setlength\LTleft{5pt}            
  
{\setlength{\tabcolsep}{4pt}
\begin{center}
\begin{longtable}{cccccccccp{3.6cm}}

\caption[List of quasar pair candidates]{Candidate quasar pairs drawn from our parent sample \protect\footnotemark.}

\label{full} \\

\hline
\hline
$\Delta\theta$ & Obs. Stat. & $\alpha$& $\delta$ & $i$  & $g$ & Classification& $z_{\rm spec}$ & QQ? \\ 
 $(\arcsec)$  & & (J2000) & (J2000) & & & & & \\
\hline
\endfirsthead

{{\bfseries \tablename\ \thetable{}  --  continued from previous page}} \\
\hline
\hline
$\Delta\theta$ & Obs. Stat. & $\alpha$& $\delta$ & $i$  & $g$ & Classification& $z_{\rm spec}$ & QQ? \\ 
$ (\arcsec)$  & & (J2000) & (J2000) & & & & & \\
\hline
\endhead

\hline \multicolumn{2}{r}{{Continued on next page.}} \\ 
\endfoot

\hline \hline

\\
\caption*{1) Angular separation of the two members of the pair; 2) The observational status of the pair is ``0" if there is insufficient information to determine the redshift of a candidate, ``1" for sources confirmed by this study, and ``2" for sources confirmed in previous studies \citep{sch05,he06a,in08,my08,og08,sch10,og12,pr13b,pr14,par16}; 3-4) Source coordinates in degrees; 5-6) dereddened $i$- and $g$-magnitudes; 7) Spectroscopic classification: Q=Quasar G=Galaxy, S=Star, U=No Spectrum, NQ=A spectrum exists but it did not yield a definitive classification (i.e ``Not a quasar"); 8) The measured or reported spectroscopic redshift for the members, -1 for objects with no redshift; 9) Classification of the pair as (1) lacking sufficient spectroscopic information to define its nature, (2) a projected pair (star-star, star-quasar, two quasars at different redshifts etc.), (3) a binary quasar, (4) a gravitational lens.}
\endlastfoot

1.449	&	2	&	184.19140	&	35.49488	&	19.88	&	20.40	&	Q	&	-1	&	4	\\	&	2	&	184.19190	&	35.49486	&	19.08	&	19.39	&	Q	&	2.013	  \\ \hline
1.693	&	2	&	204.77974	&	13.17768	&	18.91	&	19.03	&	Q	&	2.241	&	4	\\	&	2	&	204.78015	&	13.17741	&	18.87	&	18.96	&	Q	&	2.237	  \\  \hline
1.762	&	0	&	19.55013	&	-1.07848	&	19.99	&	21.00	&	U	&	0.740	&	1	\\	&	0	&	19.55053	&	-1.07820	&	20.35	&	20.40	&	U	&	-1	    \\ \hline
1.897	&	0	&	177.82866	&	46.87642	&	20.28	&	20.68	&	U	&	-1	&	1	\\	&	0	&	177.82939	&	46.87626	&	19.04	&	20.41	&	U	&	-1	    \\ \hline
1.939	&	0	&	230.20850	&	26.62804	&	19.07	&	19.35	&	Q	&	-1	&	1	\\	&	2	&	230.20911	&	26.62802	&	19.00	&	19.21	&	Q	&	1.365	     \\ \hline
1.999	&	2	&	228.91032	&	15.19300	&	18.37	&	18.70	&	Q	&	2.052	&	4	\\	&	2	&	228.91080	&	15.19331	&	18.05	&	18.16	&	Q	&	2.054	     \\ \hline
2.102	&	0	&	112.11562	&	26.11704	&	19.65	&	19.98	&	U	&	-1	&	1	\\	&	0	&	112.11615	&	26.11737	&	18.84	&	18.90	&	U	&	-1	    \\ \hline
2.173	&	0	&	146.32053	&	22.41586	&	20.79	&	20.94	&	U	&	-1	&	1	\\	&	0	&	146.32063	&	22.41646	&	20.72	&	20.81	&	U	&	-1	    \\ \hline
2.196	&	0	&	250.07547	&	10.75175	&	19.59	&	20.42	&	U	&	-1	&	1	\\	&	0	&	250.07599	&	10.75141	&	17.83	&	18.39	&	U	&	-1	    \\ \hline
2.267	&	0	&	244.24273	&	36.50716	&	20.42	&	20.17	&	U	&	-1	&	1	\\	&	0	&	244.24332	&	36.50758	&	20.40	&	20.12	&	U	&	-1	    \\ \hline
2.316	&	2	&	250.79727	&	31.93844	&	19.53	&	20.00	&	Q	&	0.587	&	1	\\	&	0	&	250.79745	&	31.93907	&	19.47	&	19.89	&	U	&	-1	    \\ \hline
2.453	&	2	&	145.64575	&	23.17533	&	19.76	&	19.91	&	Q	&	1.833	&	3	\\	&	2	&	145.64598	&	23.17468	&	19.70	&	19.80	&	Q	&	1.833	     \\ \hline
2.654	&	2	&	158.83012	&	7.88232		&	20.16	&	20.62	&	Q	&	1.218	&	3	\\	&	2	&	158.83069	&	7.88278		&	19.02	&	19.10	&	Q	&	1.215	     \\ \hline
2.678	&	0	&	161.08777	&	4.49745		&	20.60	&	20.99	&	U	&	-1	&	1	\\	&	0	&	161.08844	&	4.49713		&	19.35	&	19.73	&	U	&	-1	    \\ \hline
2.695	&	0	&	115.60587	&	24.86230	&	20.64	&	20.90	&	U	&	-1	&	1	\\	&	0	&	115.60665	&	24.86254	&	20.56	&	20.74	&	U	&	-1	    \\ \hline
2.829	&	1	&	182.49029	&	11.61649	&	20.46	&	20.76	&	Q	&	0.899	&	3	\\	&	1	&	182.49049	&	11.61573	&	20.40	&	20.65	&	Q	&	0.904	     \\ \hline
2.868	&	0	&	322.49351	&	12.00661	&	20.71	&	20.88	&	S	&	-1	&	1	\\	&	0	&	322.49390	&	12.00731	&	20.50	&	20.49	&	U	&	-1	    \\ \hline
2.903	&	2	&	227.17583	&	33.46739	&	20.60	&	20.44	&	Q	&	0.877	&	3	\\	&	1	&	227.17590	&	33.46820	&	20.56	&	20.38	&	Q	&	0.878	     \\ \hline
2.912	&	1	&	152.07303	&	17.25558	&	20.27	&	20.67	&	Q	&	1.087	&	3	\\	&	1	&	152.07367	&	17.25506	&	20.19	&	20.52	&	Q	&	1.083	     \\ \hline
2.918	&	1	&	143.15840	&	29.40301	&	20.94	&	20.66	&	U	&	-1	&	1	\\	&	0	&	143.15854	&	29.40221	&	20.90	&	20.59	&	U	&	-1	      \\ \hline
2.925	&	2	&	150.36812	&	50.46623	&	17.71	&	18.34	&	Q	&	1.845	&	4	\\	&	2	&	150.36922	&	50.46581	&	17.32	&	17.55	&	Q	&	1.841	     \\ \hline
2.933	&	1	&	158.41804	&	2.47517		&	19.65	&	20.28	&	U	&	-1	&	1	\\	&	1	&	158.41884	&	2.47531		&	19.91	&	20.14	&	Q	&	1.833	     \\ \hline
2.993	&	2	&	207.37436	&	12.45193	&	18.73	&	19.32	&	Q	&	1.722	&	4	\\	&	1	&	207.37503	&	12.45245	&	18.66	&	19.19	&	Q	&	1.722	     
\footnotetext{The full table is available in the electronic version of this paper.}

\end{longtable}
\end{center}
}

\begin{longtable}[tc]{cccccccclp{5.6cm}}

\caption[]{Complete sample of 47 spectroscopically confirmed binaries} \label{tab47} \\
  
\hline
\hline
 Name & $R.A.$& $Dec.$ & $g$  & $i$ & $\Delta\theta$& $z_{\rm spec}$ & $|\Delta v|$ & $R $\\ 
   & (J2000) & (J2000) &      &     &       $(\arcsec)$                  &     &     ($\rm km\,\rm s^{-1}$)       & $(h^{-1}$\, kpc)\\
\hline
\endfirsthead

{{\bfseries \tablename\ \thetable{}  --  continued from previous page}} \\
\hline
\hline
 Name&$R.A.$&$Dec.$& $g$  & $i$ & $\Delta\theta$& $z_{\rm spec}$ & $|\Delta v|$ & $R$\\ 
   & (J2000) & (J2000) &      &     &             $(\arcsec)$             &     &   ($\rm km\,\rm s^{-1}$)   & $(h^{-1}$\,kpc)\\

\hline
\endhead

\hline \multicolumn{2}{r}{{Continued on next page.}} \\ 
\endfoot
 
\hline \hline
\caption*{Columns: 1) Name of the members of the binary, where the brighter and fainter quasars in the pair in $g$-band are referred to as ``A'' or ``B'' respectively; 2 -- 3) Right Ascension and Declination of each quasar; 4 -- 5) $g$- and $i$-magnitude of
 each quasar; 6) Angular separation of the quasars in the binary; 7) Spectroscopic redshift for the binary; 8) The velocity difference between the quasars in the binary; 9) The transverse proper separation between the quasars in the binary.}

\endlastfoot

SDSS J0718+4020	A	&	109.51462	&	40.35075	&	20.04	&	20.16	&	5.926	&	1.838	&	0	&	34.6	\\
SDSS J0718+4020	B	&	109.51288	&	40.34978	&	20.66	&	21.07	&		&		&		&		\\
\\
SDSS J0751+1303	A	&	117.76192	&	13.06113	&	20.31	&	20.82	&	6.166	&	1.545	&	1300	&	36.1	\\
SDSS J0751+1303	B	&	117.76159	&	13.05944	&	20.38	&	20.94	&		&		&		&		\\
\\
SDSS J0813+5416	A	&	123.30461	&	54.27972	&	17.27	&	17.24	&	5.042	&	0.778	&	200	&	26.1	\\
SDSS J0813+5416	B	&	123.30266	&	54.28054	&	20.20	&	20.25	&		&		&		&		\\
\\
SDSS J0818+3623	A	&	124.63308	&	36.38616	&	17.73	&	17.99	&	6.094	&	1.961	&	0	&	35.3	\\
SDSS J0818+3623	B	&	124.63222	&	36.38770	&	19.26	&	19.75	&		&		&		&		\\
\\
SDSS J0846+2709	A	&	131.60213	&	27.16733	&	20.44	&	20.43	&	4.637	&	2.195	&	0	&	26.5	\\
SDSS J0846+2709	B	&	131.60140	&	27.16622	&	20.46	&	20.66	&		&		&		&		\\
\\
SDSS J0916+3252	A	&	139.24397	&	32.87321	&	19.38	&	19.73	&	6.122	&	1.911	&	600	&	35.6	\\
SDSS J0916+3252	B	&	139.24195	&	32.87304	&	19.75	&	20.10	&		&		&		&		\\
\\
SDSS J0922-0117	A	&	140.57307	&	-1.29715	&	18.64	&	18.93	&	6.032	&	1.677	&	1400	&	35.3	\\
SDSS J0922-0117	B	&	140.57305	&	-1.29883	&	19.48	&	19.83	&		&		&		&		\\
\\
SDSS J0954+1920	A	&	148.62408	&	19.33632	&	18.41	&	18.58	&	4.376	&	1.744	&	0	&	25.6	\\
SDSS J0954+1920	B	&	148.62282	&	19.33660	&	19.96	&	20.29	&		&		&		&		\\
\\
SDSS J0959+5449	A	&	149.78113	&	54.81844	&	19.74	&	20.04	&	3.945	&	1.956	&	200	&	22.9	\\
SDSS J0959+5449	B	&	149.77942	&	54.81892	&	20.29	&	20.61	&		&		&		&		\\
\\
SDSS J1048+0950	A	&	162.19373	&	9.83695		&	20.56	&	20.74	&	4.447	&	1.666	&	0	&	26.1	\\
SDSS J1048+0950	B	&	162.19254	&	9.83652		&	20.61	&	20.84	&		&		&		&		\\
\\
SDSS J1145+2857	A	&	176.26947	&	28.95353	&	19.97	&	20.24	&	4.085	&	2.173	&	100	&	23.4	\\
SDSS J1145+2857	B	&	176.26818	&	28.95363	&	20.55	&	20.63	&		&		&		&		\\
\\
SDSS J1145+2857	A	&	176.93478	&	33.08547	&	17.47	&	17.58	&	4.691	&	1.164	&	1000	&	26.9	\\
SDSS J1145+2857	B	&	176.93325	&	33.08565	&	20.14	&	20.18	&		&		&		&		\\
\\
SDSS J1158+1355	A	&	179.71272	&	13.92666	&	20.78	&	20.66	&	3.237	&	2.062	&	1800	&	18.7	\\
SDSS J1158+1355	B	&	179.71198	&	13.92718	&	20.85	&	20.78	&		&		&		&		\\
\\
SDSS J1207+1408	A	&	181.86359	&	14.13900	&	19.97	&	20.11	&	3.949	&	1.795	&	500	&	23.1	\\
SDSS J1207+1408	B	&	181.86292	&	14.13811	&	20.03	&	20.37	&		&		&		&		\\
\\
SDSS J1215+0225	A	&	183.94466	&	2.43279		&	19.51	&	19.69	&	5.729	&	1.445	&	0	&	33.5	\\
SDSS J1215+0225	B	&	183.94425	&	2.43125		&	19.55	&	19.77	&		&		&		&		\\
\\
SDSS J1219+2541	A	&	184.89709	&	25.68951	&	19.61	&	19.87	&	5.897	&	1.596	&	200	&	34.6	\\
SDSS J1219+2541	B	&	184.89556	&	25.68862	&	18.64	&	20.07	&		&		&		&		\\
\\
SDSS J1235+0434	A	&	188.98030	&	68.60752	&	19.51	&	19.64	&	3.513	&	1.529	&	1800	&	20.6	\\
SDSS J1235+0434	B	&	188.97826	&	68.60689	&	19.54	&	19.72	&		&		&		&		\\
\\
SDSS J1259+1241	A	&	194.98174	&	12.69828	&	19.73	&	19.99	&	3.554	&	2.180	&	900	&	20.3	\\
SDSS J1259+1241	B	&	194.98110	&	12.69752	&	19.78	&	20.09	&		&		&		&		\\
\\
SDSS J1303+5100	A	&	195.85907	&	51.01311	&	19.98	&	20.33	&	3.806	&	1.686	&	200	&	22.3	\\
SDSS J1303+5100	B	&	195.85893	&	51.01417	&	20.34	&	20.54	&		&		&		&		\\
\\
SDSS J1320+3056	A	&	200.09435	&	30.93842	&	19.65	&	19.90	&	4.745	&	1.597	&	500	&	27.8	\\
SDSS J1320+3056	B	&	200.09394	&	30.93969	&	19.68	&	19.94	&		&		&		&		\\
\\
SDSS J1337+6012	A	&	204.30472	&	60.20183	&	18.34	&	18.55	&	3.118	&	1.721	&	1500	&	18.2	\\
SDSS J1337+6012	B	&	204.30452	&	60.20269	&	19.70	&	20.05	&		&		&		&		\\
\\
SDSS J1339+6208	A	&	204.75824	&	62.14766	&	19.96	&	20.26	&	3.89	&	1.799	&	1800	&	22.7	\\
SDSS J1339+6208	B	&	204.75796	&	62.14659	&	20.36	&	20.90	&		&		&		&		\\
\\
SDSS J1344+1948	A	&	206.12888	&	19.81089	&	19.95	&	20.06	&	4.694	&	1.534	&	200	&	27.5	\\
SDSS J1344+1948	B	&	206.12883	&	19.80959	&	20.25	&	20.48	&		&		&		&		\\
\\
SDSS J1418+2441	A	&	214.73140	&	24.68464	&	19.83	&	20.15	&	4.504	&	0.573	&	400	&	20.5	\\
SDSS J1418+2441	B	&	214.73091	&	24.68581	&	19.87	&	20.23	&		&		&		&		\\
\\
SDSS J1426+0719	A	&	216.51802	&	7.32501		&	19.82	&	20.03	&	4.271	&	1.324	&	300	&	24.8	\\
SDSS J1426+0719	B	&	216.51778	&	7.32385		&	20.59	&	20.82	&		&		&		&		\\
\\
SDSS J1430+0714	A	&	217.51202	&	7.23648		&	19.02	&	19.39	&	5.414	&	1.245	&	1700	&	31.3	\\
SDSS J1430+0714	B	&	217.51110	&	7.23767		&	19.74	&	20.27	&		&		&		&		\\
\\
SDSS J1430+1539	A	&	217.51620	&	15.66371	&	19.76	&	19.64	&	6.265	&	0.912	&	200	&	34.0\\
SDSS J1430+1539	B	&	217.51486	&	15.66256	&	19.75	&	20.06	&		&		&		&		\\
\\
SDSS J1431+2705	A	&	217.77074	&	27.09129	&	20.10	&	20.19	&	5.913	&	2.261	&	900	&	33.6	\\
SDSS J1431+2705	B	&	217.76937	&	27.09018	&	20.13	&	20.25	&		&		&		&		\\
\\
SDSS J1433+1450	A	&	218.46286	&	14.83489	&	19.21	&	19.38	&	3.336	&	1.506	&	500	&	19.5	\\
SDSS J1433+1450	B	&	218.46227	&	14.83561	&	19.25	&	19.45	&		&		&		&		\\
\\
SDSS J1439+0601	A	&	219.95763	&	6.01756		&	20.39	&	20.80	&	5.329	&	1.151	&	0	&	30.4	\\
SDSS J1439+0601	B	&	219.95697	&	6.01889		&	20.47	&	20.94	&		&		&		&		\\
\\
SDSS J1440+1515	A	&	220.24983	&	15.26339	&	19.55	&	19.97	&	3.852	&	1.153	&	0	&	22.0	\\
SDSS J1440+1515	B	&	220.24968	&	15.26233	&	20.48	&	20.65	&		&		&		&		\\
\\
SDSS J1444+5413	A	&	221.09413	&	54.22240	&	20.05	&	20.25	&	3.446	&	1.584	&	0	&	20.2	\\
SDSS J1444+5413	B	&	221.09255	&	54.22263	&	20.15	&	20.83	&		&		&		&		\\
\\
SDSS J1457+2516	A	&	224.49562	&	25.28052	&	19.71	&	19.82	&	5.689	&	1.376	&	0	&	33.2	\\
SDSS J1457+2516	B	&	224.49422	&	25.27957	&	19.82	&	19.99	&		&		&		&		\\
\\
SDSS J1458+5448	A	&	224.61137	&	54.80367	&	19.62	&	20.49	&	5.142	&	1.905	&	300	&	29.9	\\
SDSS J1458+5448	B	&	224.60902	&	54.80413	&	20.47	&	20.80	&		&		&		&		\\
\\
SDSS J1507+2903	A	&	226.94681	&	29.05924	&	19.86	&	19.88	&	4.349	&	0.863	&	0	&	23.2	\\
SDSS J1507+2903	B	&	226.94545	&	29.05949	&	20.19	&	20.44	&		&		&		&		\\
\\
SDSS J1512+2951	A	&	228.24347	&	29.86401	&	18.38	&	18.58	&	5.312	&	1.809	&	500	&	31.0	\\
SDSS J1512+2951	B	&	228.24194	&	29.86337	&	19.52	&	20.83	&		&		&		&		\\
\\
SDSS J1518+2959	A	&	229.59763	&	29.99099	&	19.87	&	20.17	&	5.281	&	1.249	&	800	&	30.5	\\
SDSS J1518+2959	B	&	229.59607	&	29.99042	&	19.92	&	20.25	&		&		&		&		\\
\\
SDSS J1530+5304	A	&	232.66176	&	53.06685	&	20.28	&	20.64	&	4.114	&	1.535	&	200	&	24.1	\\
SDSS J1530+5304	B	&	232.66068	&	53.06779	&	20.31	&	20.70	&		&		&		&		\\
\\
SDSS J1545+2755	A	&	236.31659	&	27.93363	&	19.43	&	19.70	&	3.735	&	1.494	&	100	&	21.9	\\
SDSS J1545+2755	B	&	236.31556	&	27.93314	&	20.24	&	20.66	&		&		&		&		\\
\\
SDSS J1553+2230	A	&	238.37730	&	22.50399	&	20.50	&	20.70	&	6.111	&	0.641	&	1300	&	29.2	\\
SDSS J1553+2230	B	&	238.37596	&	22.50284	&	20.63	&	20.93	&		&		&		&		\\
\\
SDSS J1559+2640	A	&	239.78497	&	26.67552	&	19.66	&	19.81	&	5.367	&	0.870	&	1000	&	28.7	\\
SDSS J1559+2640	B	&	239.78424	&	26.67686	&	20.29	&	20.44	&		&		&		&		\\
\\
SDSS J1602+1314	A	&	240.61542	&	13.23796	&	19.93	&	20.08	&	5.324	&	2.018	&	100	&	30.8	\\
SDSS J1602+1314	B	&	240.61448	&	13.23680	&	20.35	&	20.46	&		&		&		&		\\
\\
SDSS J1606+2900	A	&	241.51259	&	29.01413	&	18.37	&	18.29	&	3.446	&	0.770	&	300	&	17.7	\\
SDSS J1606+2900	B	&	241.51172	&	29.01355	&	18.36	&	18.50	&		&		&		&		\\
\\
SDSS J1635+2911	A	&	248.79294	&	29.18783	&	20.11	&	20.33	&	4.917	&	1.587	&	0	&	28.8	\\
SDSS J1635+2911	B	&	248.79228	&	29.18907	&	20.16	&	20.42	&		&		&		&		\\
\\
SDSS J1637+2636	A 	&	249.25389	&	26.60274	&	18.97	&	19.11	&	3.904	&	1.961	&	0	&	22.6	\\
SDSS J1637+2636	B	&	249.25367	&	26.60381	&	20.52	&	20.60	&		&		&		&		\\
\\
SDSS J1649+1733	A	&	252.37083	&	17.55239	&	19.23	&	19.45	&	3.618	&	2.080	&	0	&	20.8	\\
SDSS J1649+1733	B	&	252.36997	&	17.55182	&	19.42	&	19.77	&		&		&		&		\\
\\
SDSS J1723+5904	A	&	260.82260	&	59.07956	&	18.56	&	18.78	&	3.721	&	1.597	&	400	&	21.8	\\
SDSS J1723+5904	B	&	260.82211	&	59.07855	&	20.07	&	20.32	&		&		&		&		\\

\end{longtable}

\end{document}